\documentclass{aa}  

\usepackage{graphicx}
\usepackage{pdflscape}
\usepackage{color}
\usepackage{gensymb}
\usepackage{txfonts}
\usepackage{natbib}
\bibpunct{(}{)}{;}{a}{}{,} 

\begin{document}

\title{A Gaia DR~2 and VLT/FLAMES search for new satellites of the LMC\thanks{Based on ESO programs 096.B-0785(A) and 098.B-0419(A).}} 

\subtitle{}

   \author{T. K. Fritz
          \inst{1,2}
           \and R. Carrera\inst{3}
          \and
          G. Battaglia\inst{1,2}\and
           S. Taibi\inst{1,2}
          }

   \institute{Instituto de Astrofisica de Canarias, calle Via Lactea s/n, E-38205 La Laguna, Tenerife, Spain\\
              \email{tfritz@iac.es}
         \and
         Universidad de La Laguna, Dpto. Astrofisica, E-38206 La Laguna, Tenerife, Spain   
       \and  INAF - Osservatorio Astronomico di Padova, Vicolo dell’Osservatorio 5, I-35122 Padova, Italy 
             }

   \date{}

 
  \abstract{
  {A wealth of tiny galactic systems populates the surroundings of the Milky Way. However, some of these objects might actually have their origin as former satellites of the Magellanic Clouds, in particular of the LMC. Examples of the importance of understanding how many systems are genuine satellites of the Milky Way or the LMC are the implications that the number and luminosity/mass function of satellites around hosts of different mass have for dark matter theories and the treatment of baryonic physics in simulations of structure formation.
}
{Here we aim at deriving the bulk motions and estimates of the internal velocity dispersion and metallicity properties in four recently discovered distant southern dwarf galaxy candidates, Columba~I, Reticulum~III, Phoenix~II and Horologium~II.}
{We combine \textit{Gaia} DR2 astrometric measurements, photometry and new FLAMES/GIRAFFE intermediate resolution spectroscopic data in the region of the near-IR Ca~II triplet lines; such combination is essential for finding potential member stars in these low luminosity systems.}
{We find very likely member stars in all four satellites and are able to determine (or place limits on) the systems bulk motions and average internal properties. The systems are found to be very metal-poor, in agreement with dwarf galaxies and dwarf galaxy candidates of similar luminosity. 
}
{Among the four objects, the only one that we can place firmly in the category of dwarf galaxies is Phoenix~II given its resolved large velocity dispersion  ($9.5_{-4.4}^{+6.8}\,$km/s) and intrinsic metallicity spread (0.33$\,$dex). Also for Columba~I we measure a clear metallicity spread. The orbital pole of Phoenix~II is well constrained and close to that of the LMC, suggesting a prior association. The uncertainty on the orbital poles of the other systems are presently very large, so that an association cannot be excluded, apart from Columba~I. Using the numbers of potential former satellites of the LMC identified here and in the literature, we obtain for the LMC a dark matter mass of M$_\mathrm{200}=1.9_{-0.9}^{+1.3}\times10^{11}$\,M$_{\odot}$.
   } 
  } 

   \keywords{
Astrometry - Proper motions - Galaxies: dwarf - Galaxies: kinematics and dynamics - Local Group - Galaxies: evolution
   }

\titlerunning{A search for LMC satellites}
\authorrunning{T. Fritz et al.}   

   \maketitle
%
\section{Introduction}
\label{sec:intro}

In the $\Lambda$Cold Dark Matter ($\Lambda$CDM) framework, not only large galaxies, but also low mass halos are expected to host their own systems of satellite sub-halos. How many of these will contain a luminous component depends on several variables, among which the mass of the host halo, the mass and build-up history of the sub-halos themselves and various environmental factors, including the strength of the UV-ionizing background 
(see e.g. the review by Bullock \& Boylan-Kolchin \citeyear{Bullock_17} and references there-in). 

Observationally, there have already been several detections of low-luminosity galaxies possibly physically associated to stellar systems of similar or lower stellar mass than the LMC (e.g. Antlia~A and the recently discovered Antlia~B around NGC~3109, Sand et al. \citeyear{Sand_15}; Scl-MM-Dw1 around NGC~253, Sand et al.  \citeyear{Sand_14}); in some cases, the "status" of satellite galaxy is guaranteed by the on-going tidal disruption of such systems \citep[e.g.][]{Rich_12,Amorisco_14,Annibali_16,Toloba_16}. This could in principle be interpreted as a qualitative confirmation of one of the predictions of the $\Lambda$CDM hierarchical formation framework.

Recently, about two dozens low-luminosity, candidate dwarf galaxies were discovered at projected locations in the sky close to the Magellanic Clouds \citep{Drlica-Wagner_15,Koposov_15a,Martin_15,Bechtol_15,Laevens_15,Drlica-Wagner_16,Torrealba_18,Kim_15a,Kim_15b,Koposov_18}. 

This has of course raised the question of whether some of these systems might be, or were before infall, part of a satellite system of the Clouds rather than of the Milky Way (MW). Tackling how many, and which ones, of these dwarf satellites might have been brought in by the Clouds would give insights into several aspects of galaxy formation in a cosmological context: besides improving the current observational information on the properties of satellite systems around galaxies of lower mass than the MW, it would allow to make further considerations into the efficiency of galaxy formation at the low-mass end \citep[see e.g.][]{Dooley_17}, and might imply a revision of our understanding of the properties of the MW satellite system itself, in terms of the number of its members, as well as its luminosity and circular velocity function. Identifying {\it which ones} of these dwarf galaxies in particular might be/have been associated to the Clouds gives also a direct avenue to start addressing the effects of group pre-processing onto the observed properties of dwarf galaxies from the detailed perspective of resolved stellar population studies. 

There have been predictions on the number of satellites that could be associated to systems with stellar masses similar to the Magellanic Clouds, and in what stellar mass range they should be found \citep[see e.g.][]{Dooley_17}. A conclusion from the aforementioned study is that there is a dearth of ``massive'' satellites around the LMC and SMC; this could imply a Magellanic-Clouds ``missing satellite problem'', although other solutions are possible, such as strong modifications to abundance-matching relations \citep[which at the low mass end are very uncertain, see e.g.][and references therein]{Garrison_17,Revaz_18}, strong tidal-stripping etc.  

Several studies have instead focused on predicting which ones of the dwarf galaxies found in the surroundings of the Milky Way could have been brought in by the Magellanic system \citep{Sales_11, Deason_15, Yozin_15, Jethwa_16, Sales_17}. 

Among these, \citet{Deason_15} used the ELVIS N-body simulations to identify LMC-mass sub-haloes of MW/M31-like systems (considering virial masses in the range 1-3$\times 10^{11}$\,M$_{\odot}$) and showed that the system of satellites disperse rapidly in phase-space, unless the group has infallen recently. The sample of 25 LMC-analogs included three dynamical analogs (with similar radial and tangential velocity as observed for the LMC): the expectations in these cases are that the sub-halos found at $z=0$ within $\sim$50\,kpc, or with a 3D velocity differing less $\sim$50\,km/s, from the original host have more than 50\% chance to have been part of a LMC-mass group. Nonetheless, for the whole sample considered together, systems within 50\,kpc {\it and} 50\,km/s of a LMC-mass dwarf at $z=0$ would have $>$90\% probability of having been former group members. 

\citet{Sales_17} used the LMC-analog identified in \citet{Sales_11} in the Aquarius simulations, which has a pericenter and velocity in good agreement with the measurements, to test whether any of the 20 dwarfs known at the time in the vicinity of the LMC/SMC are/were associated to the Clouds. Among the systems that had no kinematic information, they found Hor~II, Eri~III, Ret~III, Tuc~IV, Tuc~V and Phx~II to have positions and distances consistent with a Magellanic origin, although they stressed kinematic information was needed to confirm the association. In that study several objects (among which all the classical dSphs, apart from the SMC) were excluded to have been brought in by the Clouds. However, the LMC-analog virial mass before infall (M$_{\rm 200}=3.6\times 10^{10}$\,M$_{\odot}$) is at the lower end of that expected from abundance matching relations or measurements of the LMC circular velocity \citep{Marel_14}; the satellite system of a more massive LMC-analog would probably have had an intrinsically larger velocity dispersion and more extended spatial distribution; this would likely result in a wider distribution in velocity and distance space at $z=0$ for the debris of accreted material with respect to those analyzed in that study.

\citet{Jethwa_16} have built a dynamical model of the Magellanic system, which takes into account the dynamical influence of the SMC on the group satellites orbits, the dynamical friction exerted by the LMC onto the SMC, and by the MW on the Magellanic Clouds, and a non-static MW. The LMC masses explored encompass the range of those in \citet{Deason_15} and \citet{Sales_17}. Assuming isotropy for the MW sub-halo system, they found that half of the 14 Dark Energy Survey (DES) dwarfs they were considering had a probability $>$0.7 of having belonged to the LMC.

An important, general conclusion from the above works is that knowledge of the systemic radial velocities, combined to sky position and distance, can greatly aid in the identification of previous Magellanic Clouds satellites. In particular, the most compelling evidence for association is expected to be provided by the additional information afforded by knowledge of systemic proper motions; this is due to the fact that the accreted galaxies are expected to share a similar direction of the orbital angular momentum of the LMC.  

With the second \textit{Gaia} mission \citep{2016A&A...595A...1G} data release \citep[GDR2;][]{Brown_18}, the situation has dramatically improved: not only have the accuracy of the systemic proper motions of the classical MW dwarf spheroidal galaxies (dSphs)\footnote{Even though they might simply be the same class of objects in most cases, we refer to the typically brighter, passively-evolving dwarf galaxies known prior to SDSS as ``classical'' dwarf spheroidal galaxies and to those discovered posterior to that as ultra-faint dwarfs (UFDs) when their nature as dwarf galaxy has been established or ultra-faint (UF) systems or satellites, when it is still uncertain.} been significantly improved in several cases 
\citep{Helmi_18}, but such determination has finally become possible for dozens of the ultra-faint dwarf galaxies \citep{Fritz_18b,Kallivayalil_18,Massari_18,Simon_18}, while before only Segue~1 had a systemic proper motion measurement  \citep{Fritz_18}. All of the above, being in a common, absolute reference system. 

\citet[][hereafter K18]{Kallivayalil_18} used the \citet{Sales_11, Sales_17} LMC-analog to test a possible association to the LMC for 32 UFDs with $M_V \gtrsim -8$. For the systems for which 3D velocities could be obtained, given the additional availability of published spectroscopic data, they conclude that four of those (Hor~I, Car~II, Car~III and Hyd~I) were former satellites of the Clouds, while Hyd~II and Dra~II could be reconciled with a model allowing for a larger dispersion of the tidal debris properties in velocity and distance/sky location. 

For the systems that were lacking either systemic proper motion and/or radial velocity measurements at the times of the studies, predictions are provided in several of the works cited above, under the assumption of a prior physical association to the Magellanic system. In particular, \citet{Jethwa_16} and K18 provide such predictions in the observable space of proper motion and/or radial velocity measurements, which has the advantage of not carrying the error propagation in the conversion to Galactocentric velocities. 

Here we present results from unpublished FLAMES/GIRAFFE intermediate resolution spectroscopic data available in the ESO archive for four faint and distant (70 to 190 kpc) candidate dwarf galaxies, whose location on the sky is close to the Clouds: Columba~I, Horologium~II, Phoenix~II and Reticulum~III. These were discovered in DES data \citep{Koposov_15a,Bechtol_15,Drlica-Wagner_15,Kim_15a}. Phx~II was followed up with deeper Megacam imaging \citep{Mutlu_18} and Col~I with deeper Hyper Suprime-Cam imaging \citep{Carlin_17}. We use the FLAMES/GIRAFFE data in conjunction with GDR2 astrometric information to provide the first determination of the global properties (mean spectroscopic metallicity and line-of-sight velocity dispersion) and bulk motion of these systems, and make considerations on whether these four satellites might have been former LMC satellites. 

The paper is 
structured as follows: in Sect.~\ref{sect:datasets} we describe the data-sets used for the analysis and detail the data-reduction procedure and determination of line-of-sight (l.o.s.) velocities and metallicities from the spectroscopic data in Sect.~\ref{sec:spec_ana}; in Sect.~\ref{sec:search_mem_stars} we describe the selection of stars members to the target systems; Sect.~\ref{sec:glob_prop} contains the determination of their global properties and implications for their nature as galaxies or stellar clusters. In Sect.~\ref{sec:velocities} we present the determination of space velocities and orbital properties 
of the objects in the sample and investigate a possible origin as LMC satellites. In Sect.~\ref{sec:lmcmass} we make considerations on the LMC mass suggested by the number of potential satellites, and give summary and conclusions in Sect.~\ref{sec:summ}. In the Appendix we carry out a detailed comparison to the work of \citet{Pace_18}.


\section{Data-sets}
\label{sect:datasets}

We analyze four ultra-faint (UF) systems for which public FLAMES/GIRAFFE data exist in the ESO archive (program IDs: 096.B-0785 and 098.B-0419) with no associated publication to this date to the best of our knowledge: Columba~I (Col~I), Phoenix~II (Phx~II), Horologium~II (Horo~II) and Reticulum~III (Ret~III). We complement the spectroscopic data with GDR2 \citep{Prusti_16,Brown_18} and public DECam photometry of individual point-sources from the
 NOAO source catalog first data release \citep{Nidever2018}. The latter use  dark energy survey (DES) images. 

We use the following distance moduli in the analysis: 21.31$\pm$0.11 for Col~I \citep{Carlin_17}, 19.46$\pm$0.20 for Horo~II \citep{Kim_15a}, 19.60$\pm$0.10 for Phx~II \citep{Mutlu_18} and 19.82$\pm0.31$ for Ret~III \citep{Drlica-Wagner_15}. We also add in quadrature an additional error of 0.1 mag as saveguard against systematic errors, as in \citet{Fritz_18b}.

\section{Spectroscopic analysis}
\label{sec:spec_ana}

The observations were carried out with FLAMES mounted at Very Large Telescope (VLT) UT2 \citep{2002Msngr.110....1P}, together with the GIRAFFE spectrograph and the LR8 grating, which provides a resolution of $\sim$6\,500 in the region of the near-infrared \ion{Ca}{II} triplet lines around 8\,500\,\AA. 
The data consist of 3x3000s exposures and one of 1980s for Phx~II, 6$\times$2775\,s for Col~I, 3$\times$2775\,s for Horo~II and 3$\times$2775\,s for Ret~III.

\subsection{Data reduction and extraction of spectra}
\label{sec:spectra_extrac}

\begin{figure}
   \centering
   \includegraphics[width=\columnwidth]{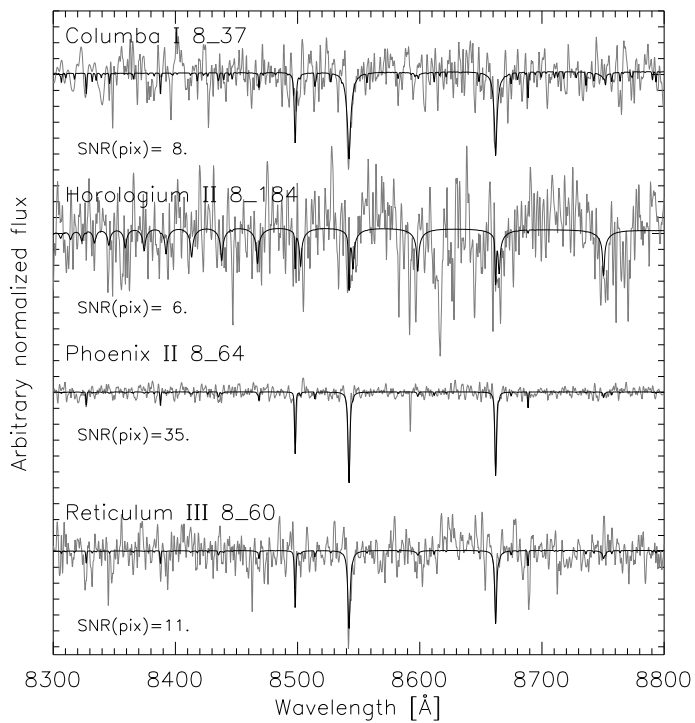}
      \caption{Example of observed spectra (grey) with different SNR and the best synthetic models (black) found in the l.o.s. velocity determination analysis. Note that Horologium~II 8\_184 is a hot star, likely a blue horizontal branch (BHB) star, and the continuum normalization is not optimized for these objects. 
     }
         \label{fig_fit}
   \end{figure}

The FLAMES/GIRAFFE data have been processed through the GIRAFFE data-reduction pipeline \citep{Melo_09}. This pipeline performs the bias, flat field and scattered light corrections; finds, traces, and extracts the spectra; and obtains the wavelength calibration based on daytime calibration exposures. Our own software was used to subtract the sky emission following the procedure described in detail by \citet[][see also Battaglia et al. 2008]{2017MNRAS.470.4285C}. Briefly, a master sky is obtained by averaging all the spectra obtained from fibers located on sky positions. The resulting master sky spectrum and the spectra for each object are separated into two components: continuum and line. To obtain the continuum of both sky and object lines we used a nonlinear median filter with 3-$\sigma$ clipping. The line
spectrum is obtained by subtracting the continuum. The sky- and object-line
components are compared to search for the scale factor
that minimizes the sky line residuals. In practice, this optimum scaling factor is the
value that minimizes the sum of the absolute differences
between the object-line and the sky-line multiplied by the
scale factor, known as L1 norm. The object-continuum is added back to the sky subtracted
object-line spectrum. Finally, the sky continuum
is subtracted assuming that the scale factor is the same as
for the sky-line component.

After applying the barycentric correction, individual spectra were averaged to obtain the combined spectrum for each star using the individual signal-to-noise ratio (SNR) as weight and an average sigma clipping rejection algorithm to remove deviant pixels. Those individual exposures with very low SNR, i.e. SNR $<$3\,/pixel, are rejected. 
This ``combined'' spectrum is then cross correlated with each individual exposure to remove small shifts between them. This procedure is repeated until convergence.

\subsection{Line-of-sight velocity and [Fe/H] determination}
\label{sec:spec_star_prop}
At this point, the ``combined'' spectrum is cross correlated with a grid of 432 synthetic spectra. The details about the computation of this grid can be found in \citet{allendegrids}. The grid has three dimensions: metallicity, [Fe/H]; effective temperature, $T_{\rm eff}$; and surface gravity, log $g$. Metallicity ranges from -5.0 to +1.0\,dex with a step of 0.5\,dex. Temperature goes from 3\,500 to 6\,000\,K with a step of 500\,K. Finally, the gravity covers from log$g=$ 0.0 to 5.0\,dex with a step of 1.0\, dex. For the $\alpha$-elements abundances the spectra were computed assuming [$\alpha$/Fe]=0.5\,dex for [Fe/H]$\leq-1.5$\,dex, [$\alpha$/Fe]=0.0\,dex for [Fe/H]$\geq$+0.0\,dex, and linear between them. The abundances of other elements were fixed to the Solar values \citep{asplund2005} and the microturbulence velocity was fixed to 1.5\,km\,s$^{-1}$. First, each spectrum is cross-correlated with a reference synthetic spectrum, which has the Arcturus parameters, to obtain an initial shift. After applying this initial shift, the observed spectrum is compared with the whole grid in order to identify the model parameters that best reproduces it through a $\chi^2$ minimization using FER\reflectbox{R}E\footnote{Available at https://github.com/callendeprieto/ferre} \citep{2006ApJ...636..804A}. The best-fitting synthetic spectrum is cross-correlated again with the observed spectrum in order to refine the shift between both. Of course, Arcturus is not the ideal template for some of the target stars. However, after the first initial determination of the shift with Arcturus, our procedure converges by itself to appropriate templates for these objects. There are a few stars for which this procedure yields temperatures close to the edge of the synthetic spectra grid, 6\,000\,K, suggesting that they may be hotter. For these particular cases, we repeated the procedure using another synthetic spectrum grid covering a temperature range between 5\,500 and 8\,500\,K also with a step of 500\,K. The other features of this grid are the same that the previous one except for the log$g$, which starts at 1.0\,dex. An example of the obtained fits is shown in Figure~\ref{fig_fit}.

   \begin{figure}
   \centering
   \includegraphics[width=\columnwidth]{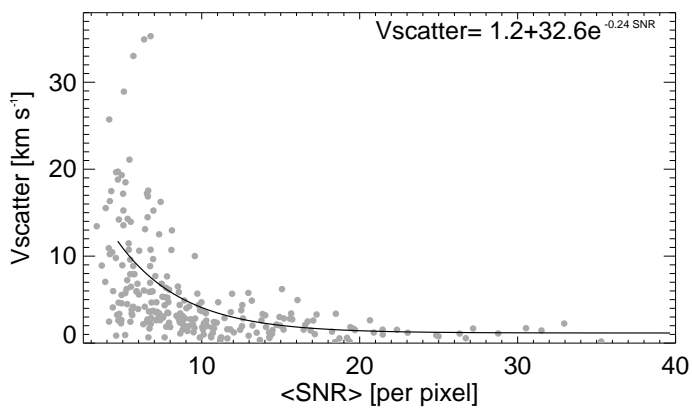}
      \caption{Run of {\sl Vscatter} as a function of median SNR of the individual exposures for all observed stars with at least 3 individual measurements. An exponential function (solid line)  has been fitted to parameterize the distribution. 
     }
         \label{fig_snr_vscatter}
   \end{figure}

The shifts obtained for each individual exposure are applied to the ``combined'' l.o.s. velocity to obtain individual values for each exposure. The final heliocentric l.o.s. velocity is obtained as the average of these  using the individual SNR as weight. The  procedure followed allows also to determine the scatter between the multiple individual l.o.s. velocity determinations, {\sl Vscatter} (see Fig.~\ref{fig_snr_vscatter}). This provides a better estimate of the internal precision than the typical uncertainty determined from the cross-correlation peak. 
For the {\sl Vscatter} calculation we excluded the short Phx~II exposure since it has clearly a lower SNR than the rest. It has to be noted that the procedure can produce low {\sl Vscatter} values by chance even at low SNR. Therefore, to better constrain the run of {\sl Vscatter} as a function of SNR we fit an exponential function to the individual {\sl Vscatter} values. According to this, the typical uncertainty at $<SNR_i>\sim$10\,/pixel is $\sim$4\,km/s.

The procedure followed to derive the l.o.s. velocities could potentially provide valuable information about the temperature, gravity, and metallicity of the observed stars. Unfortunately, the relatively small wavelength range covered, $\sim$1\,000\,\AA\@ between 8200 and 9200\,\AA, and the spectral resolution, $\sim$6\,500, do not allow to properly constrain the temperature and gravity of the observed stars, which are however necessary for the determination of the stars's metallicity. Therefore, we use photometric information to constrain the stars's temperature and gravity. To that aim, we first  transformed the \textit{Gaia} magnitudes $G$, $G_{BP}$, and $G_{RP}$ into $V$, $I$, $J$, $H$, and $K_s$ using the relationships provided by \citet{2018arXiv180409368E}. The absolute magnitudes in these bandpasses have been obtained using the distance  moduli listed at the beginning of Sect.~\ref{sect:datasets}. Temperatures have been derived using the relationships by \citet{2005ApJ...626..465R} assuming an initial estimation of metallicity of [M/H]$=-2.0$\,dex. We derived temperatures for $V-I$, $V-J$, $V-H$, and $V-K_s$ independently and have obtained the corresponding uncertainties propagating errors along the whole procedure. The main sources of uncertainty are the relationships to transform magnitudes from the \textit{Gaia} photometric system and the uncertainty in the distance. All together, our temperatures have a typical uncertainty of 100\,K.  For the hottest stars the uncertainties are slightly higher.  Since the values obtained from the different colors agree within the uncertainties, we obtain the final temperature by averaging them together.

Surface gravities have been determined using the Bayesian estimation algorithm\footnote{Available at \url{http://stev.oapd.inaf.it/cgi-bin/param_1.3}} presented by \citet{2006A&A...458..609D} with PAdova and TRieste Stellar Evolution Code (PARSEC) evolutionary tracks \citep{2012MNRAS.427..127B} and the temperatures above. The average uncertainty in the $\log g$ determination is 0.15\,dex. Finally, the metallicity has been determined from the combined spectra by comparing the observed spectra with the same synthetic grid used in the l.o.s. velocity determination but fixing the temperatures and gravities to the photometric values. Our analysis confirms that one of the observed stars, Horo~II 8\_184 is likely a blue horizontal branch star as expected from its position on the color-magnitude diagram (see Figures~\ref{vhelio}). 

\begin{figure}
   \centering
   \includegraphics[width=\columnwidth]{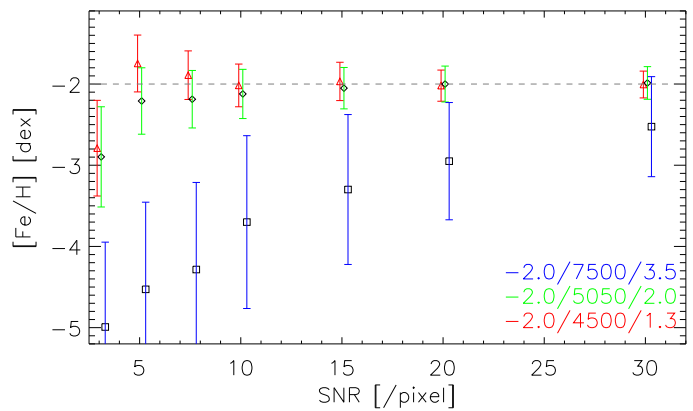}
      \caption{Recovered metallicities as a function of SNR for three different synthetic spectra with the input stellar parameters listed on bottom-right corner ([Fe/H]/effective temperature/log\,$g$). Note that the SNR for each model has been slightly shifted to avoid superposition of the symbols. }
         \label{fig_test_metal}
   \end{figure}

Owing to several of our observed stars having a low SNR, it is necessary to check the reliability of our results. To that aim, we selected synthetic models with metallicity [Fe/H]=-2.0\,dex but with three different temperatures (4\,500, 5\,050, and 7\,500\,K) and surface gravities (1.3, 2.0, and 3.5\,dex), which covers the whole range of temperatures and gravities of the stars in our observed sample. Different levels of noise have been injected to each spectrum in order to obtain spectra of different SNRs. Fixing the temperature and gravity to their input values, the spectra have been analyzed as the observed stars in order to determine their metallicity as a function of the SNR (Fig.~\ref{fig_test_metal}). At SNR=3\,/pixel, the recovered metallicity is underestimated with respect to the input one for the three models. At SNR\,$>$5\,/pixel the input metallicity is recovered within the uncertainties for the two coldest models. In fact, for the coldest model the method works slightly better than for the other one at a given SNR. For the hottest model, which represents the blue horizontal branch star in Horo~II (see Sect.~\ref{sec:search_mem_stars}), the recovered [Fe/H] is always underestimated, although the difference with the input value decreases as SNR increases.

      \begin{table*}
      \small
      \caption[]{Number of stars targeted in the FLAMES/GIRAFFE observations, retained after successive cleaning criteria; rows 4-5-6-7 apply only to stars with GDR2 kinematic information, whilst rows 8-9-10 to those without. From the top to the bottom row: number of stars with FLAMES/GIRAFFE spectroscopic observations; stars for which the radial velocities have been determined;
      retained after the photometry-based cleaning; after parallax cut; after proper motion cut. Rows 6 and 7 give the number of stars that we consider as certain or candidate members, respectively, when also the l.o.s. velocity is used as criterion. Row 8 lists the number of stars w/o GDR2 kinematic information; the last two rows give how many of them are certain and candidate members when the l.o.s velocity information is included.}
         \label{KapSou0}
      $$
         \begin{array}{p{0.19\linewidth}llll}
\hline
\hline
  method &  \mathrm{Columba~I} & \mathrm{Horologium~II} & {\rm Phoenix~II} & \mathrm{Reticulum~III} \\
            \hline            
all spectra &    76	 &  113	 &  105	&    75	\\
good spectra  &  49	 &   93	 &   77	 &   45	\\
photometric  &  49	&    23	 &   20	 &   43	\\
\hline
parallax  &  36	  &  15	 &   15	 &   28	\\
proper motion  &  14	&     4	 &    7	 &   10	\\
members  &   3	  &   2	 &    5	 &    3	\\
candidates  &   3	 &    1	   &  1	 &    0	\\
\hline
w/o kinematics  &   7	 &    1	  &   0	 &    2	\\
w/o kin. members  &   2	  &   0	 &    0	 &    0	\\
w/o kin. candidates  &   1	  &   0	 &    0	 &    0  \\
    \noalign{\smallskip}
           \hline
       \end{array}
       $$
       \normalsize
  \end{table*}

      \begin{table*}
      \small
      \caption[]{
      Observed properties for potential member stars. Column 1 lists the star name, col. 2 and 3 its position on the sky, col.4 the distance from the satellite center in units of projected half light radii, col. 5 and 6 give  extinction corrected NSC DR1 DECam magnitudes in g- and i-bands, col. 7 the SNR of the combined spectrum, col. 8 and 9 the heliocentric velocity and metallicity derived from the FLAMES spectra, col. 10 and 11 the measured proper motions from GDR2 data, col. 12 whether the star is considered as very likely (1) or possible member (0.5). The non members are only shown in the electronic table.}
         \label{KapSou}
      $$
         \begin{array}{p{0.09\linewidth}lllllllccccl}
\hline
\hline
  name &  \mathrm{R.A.} & \mathrm{Dec.} & R/R_{\rm half} & m_g & m_i & \mathrm{SNR} & v_\mathrm{helio} & \mathrm{[Fe/H]} & \mu_{\alpha\,*} \mathrm{[mas/yr]} & \mu_{\delta} \mathrm{[mas/yr]} & mem \\
            \hline   
col1\_8\_16 & 82.82425 & -28.05393 & 1.25 &  21.31 & 20.2 & 4.6 & 149\pm12.1 & -1.92\pm0.2 & &  & 0.5 \\
col1\_8\_27 & 82.82958 & -28.03342 & 0.85 & 20.19 & 19.12  & 7.2 & 146.4\pm7.0 & -3.36\pm 0.15 & 0.00 \pm0.46 & -0.59\pm0.63 & 1 \\
col1\_8\_29 & 82.84424 & -28.03014 & 0.44 &  21.71 & 20.83 & 3.9 & 160.7\pm14.9 & -2.03\pm0.28 & &  & 1 \\
col1\_8\_32 & 82.85303 & -28.0272 & 0.21 &  21.55 & 20.65 & 4.1 & 161.6\pm13.4 & -2.19\pm0.23 & &  & 1 \\
col1\_8\_34 & 82.87403 & -28.03011 & 0.39 &  20.87 & 19.81 & 5.7 & 153.5\pm9.6 & -2\pm0.15 & -0.32\pm0.87 & 0.23\pm0.99 & 1 \\
col1\_8\_37 & 82.82174 & -27.99583 & 1.52 &  21.14 & 20.17 & 5.0 & 135.1\pm10.9 & -1.65\pm0.15 & -0.23\pm1.1 & 0.8\pm1.34 & 0.5 \\
col1\_8\_61 & 82.98112 & -28.08375 & 3.78 &  21.15 & 20.44 & 4.3 & 132.5\pm12.9 & -2.11\pm0.2 & 0.59\pm1.35 & 0.02\pm1.91 & 0.5 \\
col1\_8\_66 & 82.87996 & -28.03901 & 0.62 &  19.77 & 18.57 & 9.0 & 154.9\pm5.0 & -2.22\pm0.15 & 0.41\pm0.3 & -0.49\pm0.42 & 1 \\
col1\_8\_75 & 82.98464 & -27.97531 & 3.88 &  20.96 & 19.97 & 5.0 & 137.8\pm11.0 & -1.69\pm0.15 & 0.85\pm0.92 & -1.62\pm0.94 & 0.5 \\
horo2\_2\_48 & 49.33784 & -50.06129 & 4.36 &  19.14 & 18.48 & 7.8 & 163.4\pm6.2 & -1.81\pm0.15 & 1.7\pm0.27 & -0.47\pm0.42 & 0.5 \\
horo2\_8\_156 & 49.13646 & -50.02149 & 0.13 &  21.34 & 20.67 & 4.7 & 189.5\pm11.8 & -1.9\pm0.26 & 4.65\pm2.94 & 1.48\pm4.07 & 1 \\
horo2\_8\_184 & 49.06396 & -50.03182 & 2.07 &  19.88 & 20.15 & 4.9 & 158.3\pm11.3 & -3.68\pm2.45 & 0.12\pm0.7 & -0.65\pm1.12 & 1 \\
phx2\_5\_46 & 354.97731 & -54.40519 & 0.57 &  19.46 & 18.45 & 10.1 & 39\pm4.1 & -2.65\pm0.15 & 0.54\pm0.28 & -1.69\pm0.33 & 1 \\
phx2\_8\_127 & 355.03286 & -54.3918 & 1.78 &  19.43 & 18.65 & 10.3 & 43.7\pm4.0 & -2.89\pm0.15 & 0.59\pm0.29 & -1.03\pm0.36 & 1 \\
phx2\_8\_141 & 355.00857 & -54.37397 & 1.71 &  21.2 & 20.49 & 5.8 & 17.3\pm9.3 & -2.4\pm0.15 & -1.99\pm1.43 & 0.26\pm2.19 & 1 \\
phx2\_8\_24 & 354.92969 & -54.42286 & 2.72 &  20.17 & 19.95 & 5.5 & 26\pm9.9 & -1.1\pm0.18 & -0.24\pm0.86 & -1.9\pm0.9 & 0.5 \\
phx2\_8\_27 & 354.94749 & -54.41276 & 1.82 &  21.4 & 20.62 & 4.7 & 25.2\pm11.7 & -2\pm0.23 & 0.63\pm2.26 & -1.96\pm2.66 & 1 \\
phx2\_8\_64 & 354.98249 & -54.36899 & 1.54 &  18.59 & 17.5 & 15.7 & 28.9\pm3.6 & -2.45\pm0.15 & 0.49\pm0.15 & -1.06\pm0.17 & 1 \\
ret3\_2\_70 & 56.24722 & -60.42421 & 1.53 &  20.61 & 20.19 & 5.2 & 274.3\pm10.6 & -2.94\pm0.45 & 3.34\pm1.72 & -1.51\pm1.95 & 1 \\
ret3\_8\_60 & 56.39021 & -60.44914 & 0.37 &  20.48 & 19.68 & 6.4 & 273.3\pm8.2 & -2.32\pm0.15 & -0.78\pm0.89 & -1.05\pm1.12 & 1 \\
ret3\_8\_61 & 56.36044 & -60.45228 & 0.06 &  20.09 & 19.23 & 7.7 & 275\pm6.3 & -3.24\pm0.15 & -0.78\pm0.72 & 0.3\pm0.83 & 1 \\  
    \noalign{\smallskip}
           \hline
       \end{array}
       $$
       \normalsize
  \end{table*}

\section{Search for member stars}
\label{sec:search_mem_stars}

\subsection{Detection of the UF systems}
\label{sec:det_uf}
In order to select potential member stars and detect the UF system in l.o.s. velocity, we use a set of criteria making use of the photometric, astrometric and spectroscopic information to progressively clean the sample (see Table~\ref{KapSou0}  for the impact of each step on the selection).

We start by considering all the stars with a reliable l.o.s. velocity determination from the FLAMES/GIRAFFE spectra. We perform a broad selection on the color-magnitude diagram (CMD) around the locus expected for red giant branch (RGB) and horizontal branch (HB) stars by using the DECam i magnitudes and the $g-i$ color. Both are corrected for extinction using E($B-V$) from \citet{Schlegel_98} and the band coefficient from \citet{Abbott_18}. This retains nearly all FLAMES targets for Col~I and Ret~III, because they were originally observed with a similar selection. In the case of Horo~II and Phx~II, the original selection did not appear to include a color cut, therefore the number of stars is clearly reduced. 

\begin{figure*}
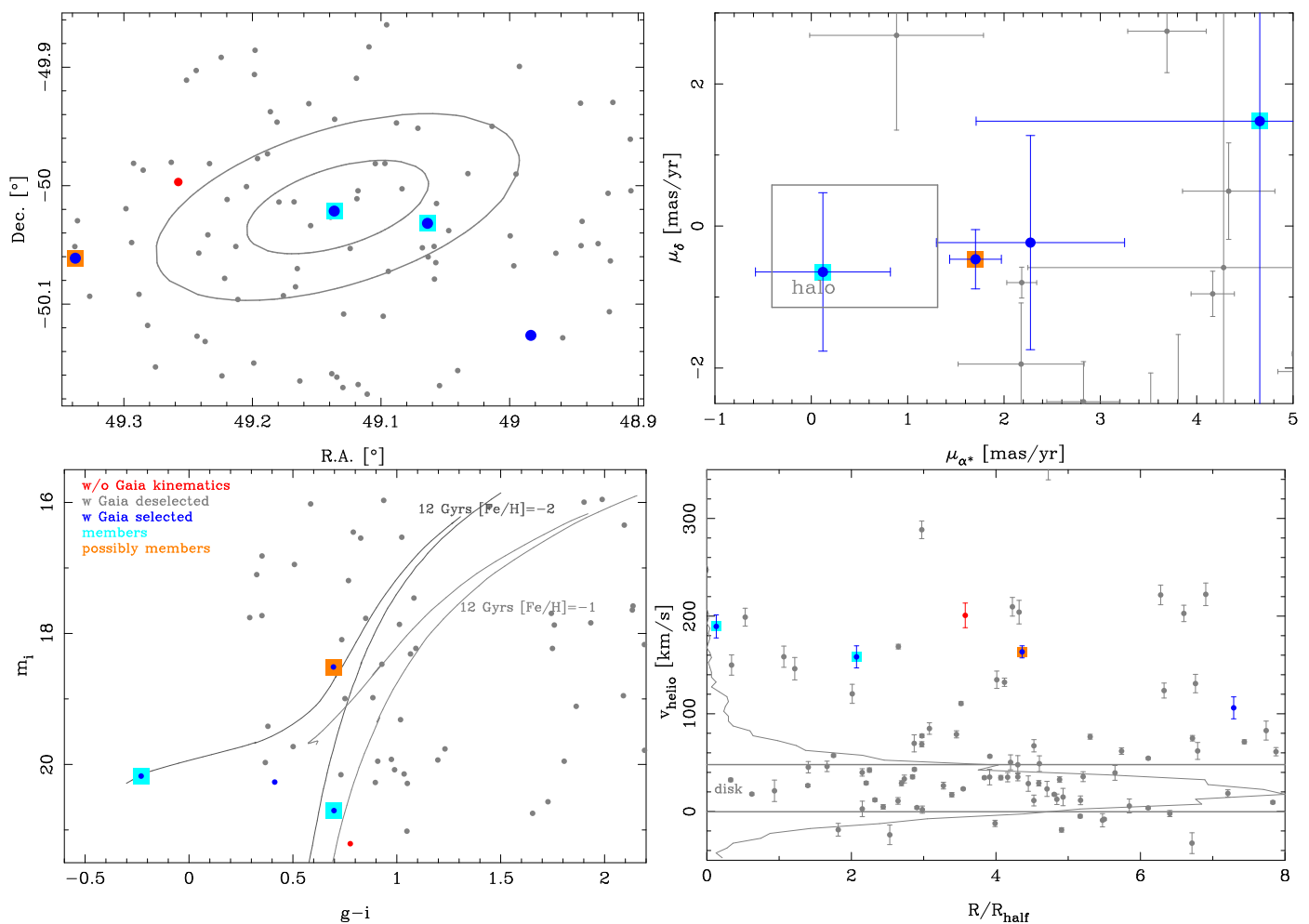

   \centering
   \includegraphics[width=0.72\columnwidth,angle=-90]{sky4_HorologiumII.eps}   
   \includegraphics[width=0.72\columnwidth,angle=-90]{mu_mu5_HorologiumII.eps}   
   \includegraphics[width=0.72\columnwidth,angle=-90]{CMD6_HorologiumII.eps}   
   \includegraphics[width=0.72\columnwidth,angle=-90]{rHelio4_HorologiumII.eps}    
      \caption{Plots for the identification of the Horologium~II members for targets with FLAMES spectroscopic observations. Stars consistent with a distant halo origin in \textit{Gaia} parallax, proper motions, and color magnitude space are plotted in blue, the not  halo stars are plotted in gray. Stars without \textit{Gaia} proper motion information, but with location on the CMD consistent with being a RGB or a BHB star at the distance of the object are plotted in red. The very likely members of the  satellite are enclosed within a cyan square, while the additional potential members within an orange square.  Top left: spatial distribution of the targets. The ellipses have semi-major axis equal to 1.5 and 3 projected (elliptical) half light radii and use ellipticity and P.A. from \citet{Kim_15b}. Top right: location in the proper motion plane; the proper motion selection box of halo stars given by the escape speed criterion is shown in gray. Bottom left: extinction corrected colour-magnitude diagram, with overlaid  Parsec isochrones with [Fe/H]$=-2.0$ and $-1.0$\,dex and age$=$\,12\,Gyrs. The isocrones have solar-scaled [$\alpha$/Fe]; 
      this might be a potential source for the discrepancy in the location of stars of a given spectroscopically determined metallicity with the isochrones of similar [Fe/H], since stars in such ancient systems are typically $\alpha$-enhanced \citep[e.g.][and references therein]{Mashonkina2017}. Bottom right: distribution on the l.o.s. $v_{\rm helio}$ versus distance from the satellite center in units of projected elliptical half-light radius. The velocity distribution of disk stars in the mock catalog ($z<|1|$\,kpc) is shown in gray, with the horizontal lines indicating the central 1\,$\sigma$ interval.
              }
         \label{vhelio}
   \end{figure*}

\begin{figure*}
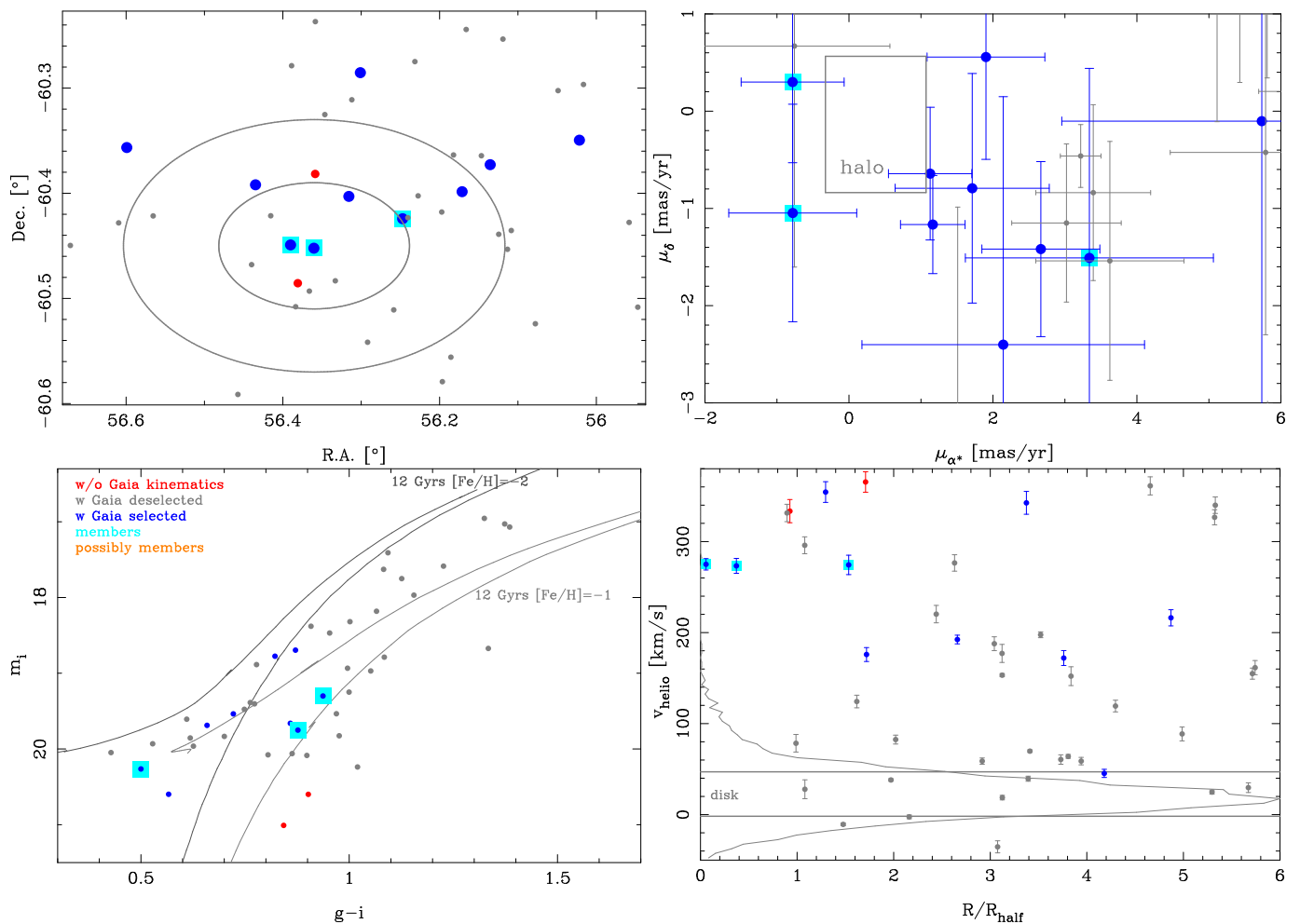

   \centering
   \includegraphics[width=0.72\columnwidth,angle=-90]{sky4_ReticulumIII.eps}   
   \includegraphics[width=0.72\columnwidth,angle=-90]{mu_mu5_ReticulumIII.eps}   
   \includegraphics[width=0.72\columnwidth,angle=-90]{CMD6_ReticulumIII.eps}   
   \includegraphics[width=0.72\columnwidth,angle=-90]{rHelio4_ReticulumIII.eps}    
      \caption{As Figure~\ref{vhelio}, but for Reticulum~III. The structural parameters are from \citet{Drlica-Wagner_15}.
              }
         \label{vhelio1}
   \end{figure*}

\begin{figure*}
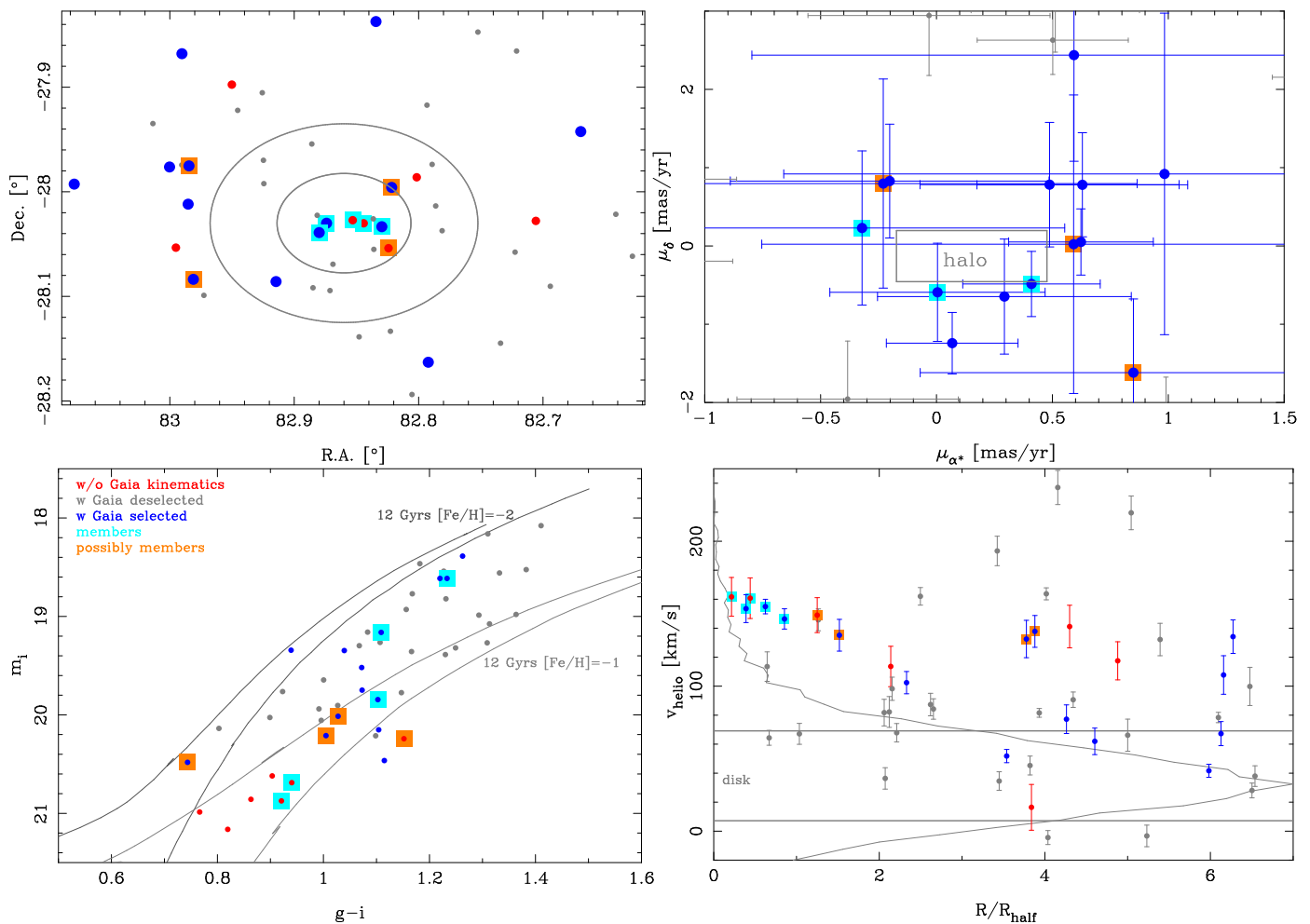

   \centering
   \includegraphics[width=0.72\columnwidth,angle=-90]{sky4_ColumbaI.eps}   
   \includegraphics[width=0.72\columnwidth,angle=-90]{mu_mu5_ColumbaI.eps}   
   \includegraphics[width=0.72\columnwidth,angle=-90]{CMD6_ColumbaI.eps}   
   \includegraphics[width=0.72\columnwidth,angle=-90]{rHelio4_ColumbaI.eps}    
      \caption{As Figure~\ref{vhelio}, but for Columba~I. The structural parameters are from \citet{Drlica-Wagner_15}.
              }
         \label{vhelio2}
   \end{figure*}

\begin{figure*}
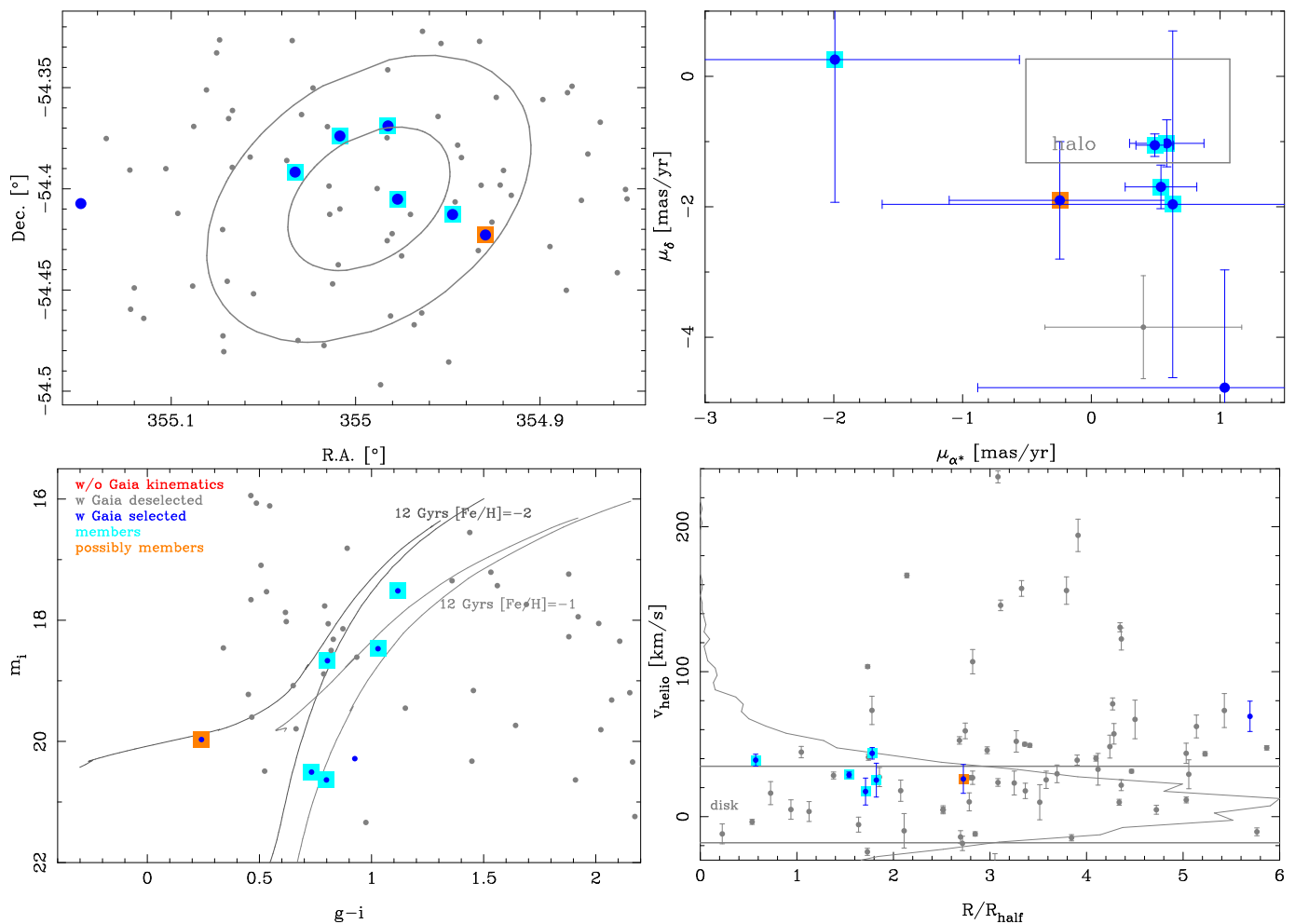

   \centering
   \includegraphics[width=0.72\columnwidth,angle=-90]{sky4_PhoenixII.eps}   
   \includegraphics[width=0.72\columnwidth,angle=-90]{mu_mu5_PhoenixII.eps}   
   \includegraphics[width=0.72\columnwidth,angle=-90]{CMD6_PhoenixII.eps}   
   \includegraphics[width=0.72\columnwidth,angle=-90]{rHelio4_PhoenixII.eps}    
      \caption{As Figure~\ref{vhelio}, but for Phoenix~II. The structural parameters are from \citet{Mutlu_18}.
              }
         \label{vhelio3}
   \end{figure*}

Then, we concentrate on those stars that have kinematic information from both GDR2 data and the FLAMES observations, because they greatly help each other to identify signals in the low number regime. We use similar criteria as in \citet{Fritz_18b} to define a sample of halo stars. Since the stars at the distances of our targets are expected to have parallaxes consistent with zero, we require that the parallax has less than 2\,$\sigma$ significance; this also removes stars with artificially very negative parallax \citep{Lindegren_18}. We further select stars by requiring them to be bound to the Milky Way given their proper motion at the distance of the satellite. We use the same generous 2\,$\sigma$ criterion as in \citet{Fritz_18b}. This selection based on parallax and proper motion greatly cleans the sample (e.g. for Ret~III the number of stars goes from 43 to 10): as it can be appreciated in Figures~\ref{vhelio}-\ref{vhelio3}, the disc component is essentially weeded out.

For those stars passing the above selection criteria, we inspect their location on the heliocentric l.o.s. velocity versus projected elliptical radius ($R$)/half-light radius ($R_{\rm half}$) plane ( blue symbols in bottom right panel of Fig.~\ref{vhelio}-\ref{vhelio3}):  in all cases some small group of stars clump at fairly similar velocity, which will be considered as our preliminary heliocentric l.o.s. systemic velocity for the satellites,
v$_\mathrm{helio, sat}$; such "velocity spike" consists of 6 stars for Col~I, 3 for Horo~II, 6 for Phx~II and  3 for Ret~III (see Sect.~\ref{sec:det_robust} for more details). 

\subsection{Testing the robustness of the detection}  
\label{sec:det_robust}

The first step is to determine whether the overall detection of a spike  in heliocentric velocity is robust for each of the UF systems analyzed.  To that aim  we determine the density of contaminant objects  at any velocity that will pass  the above CMD-based, parallax and proper motion-based selection criteria ($\rho_\mathrm{cont}$) by applying the same cuts to GDR2 sources at distances $5 < R/R_{\rm half} < 10$. As before we use stars in NSC catalogs that have measured motions in GDR2.
When determining the amount of expected contaminants, we also need to take into account the completeness of the FLAMES observations, i.e. the fact that not all potential targets had FLAMES/GIRAFFE fibers allocated on them.  For that we select from the NSC DR1 catalog potential member stars with our usual CMD selection. The faint limit is set by the faintest stars with \textit{Gaia} kinematics, while the bright limit by the brightest star targeted with  FLAMES. We also select stars in the same way from the FLAMES catalog. We then count in both catalogs how many stars are within 3 R$_\mathrm{half}$ and obtain a rough number for the completeness: $c= N_\mathrm{FLAMES}/N_\mathrm{DECam}$, which results into $c$ being between 67\% and 85\%. 

Finally,  considering $c$ and the largest radius of the potential members for each system, we translate $\rho_\mathrm{cont}$ into the number of expected contaminant stars for each satellite in the area occupied by the potential member stars ($N_\mathrm{cont}$). We obtain 2.1 for Col~I, 5.4 for Horo~II, 1.4 for Phx~II and  2.1 for Ret~III. The larger number for Horo~II is caused by a member star at 4.4R$_\mathrm{half}$ and therefore the large area considered; if this rather bright star is omitted, only 1.2 contaminants would be expected in the correspondingly smaller area occupied by the remaining member stars. Thus, in most cases the observed number of stars in the velocity spike is already somewhat larger than the expected number of contaminants, indicating a detection. 

We test how the number of expected contaminants at all velocities compares to those stars that can be considered as clear contaminants from the FLAMES observations, i.e. those falling within the radius of the outermost candidate members and passing the CMD and GDR2-based cuts, but having l.o.s. velocity outside of the velocity spike. Of these, there are two for Col~I, zero for Hor~II, zero for Phx~II, one for Ret~III. The numbers are compatible with the expectations (given small number statistics) for all cases, apart from Horo~II. The reason for this discrepancy could be partly due to the expected number of contaminants being overestimated to Horo~II because of the outermost candidate being clearly outside of the 3 half-light radii region used for the calculation of $c$.

 Could the detection of the spike be due to contaminants clustering in a given velocity range rather than to the presence of the UF system? 
In order to answer this question, we calculate what is the probability of observing a number of stars in the velocity range of the "velocity spike" of each UF system equal to or larger than the number of potential members, given a Poisson distribution with expectation value equal to $N_{\rm cont}$ and the expected $v_{\rm helio}$ distribution of contaminant stars (that, with our selection criteria, will be mostly halo stars).

Since no GDR2 l.o.s. velocity is available for sources as faint as those here studied, we  extract the information on the expected heliocentric velocity distribution of contaminants at the location of the UF systems using the 
\textit{Gaia} mock catalogue of \citet{Rybizki_18}. First, we select objects within a 1\degree~radius around the  satellites in our sample and produce new estimates of their parallax, proper motion and photometric properties by factoring in the measurement errors. Then, we apply  similar selection criteria as for the data\footnote{The only difference is that we select in \textit{Gaia} colors, since no DECam colors are given; however, we used the real data to ensure that we are being consistent with the CMD selection.}. The l.o.s. velocity distribution so obtained is broad and shows no sign of a disk population, confirming that our approach is able to select halo stars. We then make the simplifying assumption that the velocity distribution of the so-selected contaminant stars is uniform, and focus around the velocity 
of the UF system to obtain a representative value, $f_\mathrm{hel}$, which can be used as the probability of observing a star at a given velocity\footnote{In practise, we model with a quadratic function the heliocentric velocity distribution of contaminants within $\pm$50 km/s from the UF system preliminary velocity and take as representative $f_\mathrm{10\,km/s}$ that given by the ratio of the number of contaminants within $\pm$5km/s from the UF system velocity over the total. $f_\mathrm{10\,km/s}$ is then scaled to the $f_\mathrm{hel}$ of the satellite by multiplying by the velocity range of the satellite divided by 10 km/s.}. From this we finally calculate what is the probability of observing a number of contaminants within the velocity range of the "spike" equal or larger to that of the potential member stars, also accounting for that fact that we could have observed the "spike" at any velocity (i.e. we multiply the values obtained by the approximate velocity range covered by the spectroscopically observed stars, 300\,km/s, and divide by the velocity range covered by stars in the "spikes"). For all satellites, the likelihood to observe a spike of those numbers by chance is less than 0.1\%, apart from Horo~II, for which it is anyway small (2.4\%).

The realness of the spike of Horo~II could be doubted because the two potential member stars are separated by 31.2\,km/s or 2\,$\sigma$.
However, one of the two stars has the color and magnitude of an BHB star exactly at the distance of the satellite (80\,kpc). At that distance, the density of stars with BHB-like colors (which could also be blue straggler stars) is very low, about 
0.05 stars per square degree per 0.2\,mag bin, when QSOs and White dwarfs are excluded \citep{Deason_14}. The BHB candidate is found in the inner 0.017 square degree around the satellite position, thus there is less than 0.1\% probability that the star is not associated to Horo~II. Secondly, there is another brighter star within 5\,km/s of the velocity of the BHB star at 4.4 half light radii. Due to its brightness there are only 0.5 stars expected even in this correspondingly larger area. Since now the velocity of system is already known even for a single star the probability for a chance association at this $v_\mathrm{helio}$ is slightly less than 1\%. Therefore we include this star for our estimates of the mean properties of Horo~II, although we also provide values when excluding this object. The star also strengthens the detection of the systemic velocity of Horo~II. 

\subsection{Membership}
\label{sec:membership}

Now we concentrate on the different question of whether a certain star may be considered as a contaminant or a member of the satellite. We again use the photometric data at large radii (split into different magnitude bins, to account for different numbers of contaminants as a function of magnitude, i.e. $m_G<$19, 19 $< m_G <$20 and m$_G>20$) and the model for the heliocentric l.o.s. velocity distribution.
We calculate the expected background level for each candidate star with the following formula: $$n_\mathrm{cont, i}=f_\mathrm{hel}\,\rho_\mathrm{cont}\, \pi\,R_\mathrm{star}^2\,/10\sqrt{[v_\mathrm{hel,star, i}-v_\mathrm{hel, sat}]^2+3^2}.$$
The 3 (km/s) accounts for the uncertainty in the value of the observed heliocentric velocity of the satellite, while the 10 (km/s) removes the velocity range previously used to determine f$_{\rm hel}$. We use $n_\mathrm{cont, i}$ to classify stars:  those with $n_\mathrm{cont, i} < 0.1$ are treated as very likely members, while those with $n_\mathrm{cont, i}$  between 0.1 and 1 are added as potential members.  
   
The star located at 4.4 R$_\mathrm{half}$ in Horo~II remains classified only as candidate member. A star in Phx~II at 2.8 R$_\mathrm{half}$ is formally a member with n$_\mathrm{cont, i}=0.085$ but has a unusually high metallicity (see Tab.~\ref{KapSou}), therefore we consider it as candidate member, rather than a very likely member. 

In total we  find 13 very likely members (3 in Col~I, 2 in Horo~II, 5 in Phx~II, 3 in Ret~III), which are contained within 2.1 R$_\mathrm{half}$ (without the aforementioned BHB star, the largest distance would be at 1.8 R$_\mathrm{half}$). In addition we have 5 candidates: 3 for Col~I, 1 for Horo~II and 1 for Phx~II, usually at larger distances.

As a further step, we also consider the stars with FLAMES spectroscopic measurements but without \textit{Gaia} kinematic information (see red symbols in Figs.~3-6). Only Col~I has clearly promising candidates, therefore we ignore the other satellites. We use the NSC catalog to determine the background density between 5 R$_\mathrm{half}$ and 0.5\degree~using the color box of Col~I for m$_i$ between 20.4 and 21.5 and then use the l.o.s. velocity information as before: 
the two innermost stars have less than 3\% probability to be contaminants and thus we add them to our sample of likely members. The third innermost star has a $\sim$3\% probability of being a contaminant but it is also significantly redder than the other stars in the CMD; we therefore classify it as a candidate. We should note that the inclusion or exclusion of these new members to the sample does not significantly influence the overall properties of the system (determined taking into account error-bars) given the low SNR and large measurement errors of these faint stars. In total we have 5 very likely and four possible members for Col~I.

Table~\ref{KapSou} lists the classification of a given star as a very likely member or candidate, together to its l.o.s. heliocentric velocity, [Fe/H] and, when available, the proper motion information.  We highlight the presence in Col~I of an extremely metal poor star, col1\_8\_27, with m$_i=19.1$ and [Fe/H]$=-3.36$\,dex. 

 Finally, we note that the location of the member stars on the CMD appears to be too red with respect to the location expected from PARSEC isochrones of the same metallicity as the spectroscopic values. We have checked that the same effect is visible in NSC DR1 photometry of the globular cluster NGC\,1904, and therefore it is not intrinsic to our analysis; also we note that Dartmouth isochrones of the same metallicity and age would lead to slightly redder colours than the PARSEC ones. In addition, we checked that also the DES~DR1 catalog \citep[used by][]{Pace_18} leads also to a similar offset, see also Appendix~\ref{ap_pace}. All this suggests the cause of the mismatch is more likely to be due to the isochrones rather than to the photometry.

\section{Global properties and nature of the satellites}
\label{sec:glob_prop}

We now focus on the determination of the global kinematic and metallicity properties of the objects, i.e. their systemic l.o.s. velocity <$V_\mathrm{hel}$>, systemic proper motion in right ascension (<$\mu_{\alpha^*}$>) and declination (<$\mu_{\delta}$>), the average correlation coefficient between $\mu_{\alpha*}$ and $\mu_\delta$, mean metallicity (<[Fe/H]>), as well as the intrinsic spread in the l.o.s. velocity ($\sigma_{V, \mathrm{hel}}$) and metallicity distribution ($\sigma_{\mathrm{[Fe/H]}}$) of the stars. These latter two quantities are especially important for systems whose nature is uncertain such as those in the luminosity and surface brightness regime of Milky Way UF systems, since the current thinking is that large dynamical mass-to-light ratios (as revealed by an intrinsic velocity dispersion significantly larger than accounted for by the baryonic component) and/or an intrinsic spread in metallicity can be used to classify the system as a galaxy \citep[or defined to be, see ][]{Willman_12}.

Given the large uncertainties in the individual proper motion measurements, we do not attempt to derive the intrinsic spread in the distribution of the proper motion values, and we perform a simple weighted average to determine <$\mu_{\alpha^*}$>, <$\mu_{\delta}$> and the correlation coefficient. On the other hand, we adopt a Bayesian approach for the determination of the other quantities, as in \citet{Taibi_18} by running the MultiNest code \citep{Feroz2009,Buchner2014}, a multi-modal nested sampling algorithm. The nested sampling \citep{Skilling2006} is a Monte Carlo method aimed at calculating efficiently the Bayesian evidence, providing as a by-product the posterior parameter estimation. We make the assumption that the metallicity and l.o.s. velocity distributions are Gaussian and take into account the measurement errors on the individual quantities in the determination of the intrinsic spread. We assume flat priors between 0 and 100 km/s for $\sigma_{V, \mathrm{hel}}$ and between 0 and 3\,dex for $\sigma_{\mathrm{[Fe/H]}}$. Table~\ref{KapSou2} summarizes the results both for the sample of very likely members only, and for the sample that includes also potential members (we also highlight our preferred cases, as in the explanations below).

\begin{figure*}
   \centering
   \includegraphics[width=0.99\columnwidth]{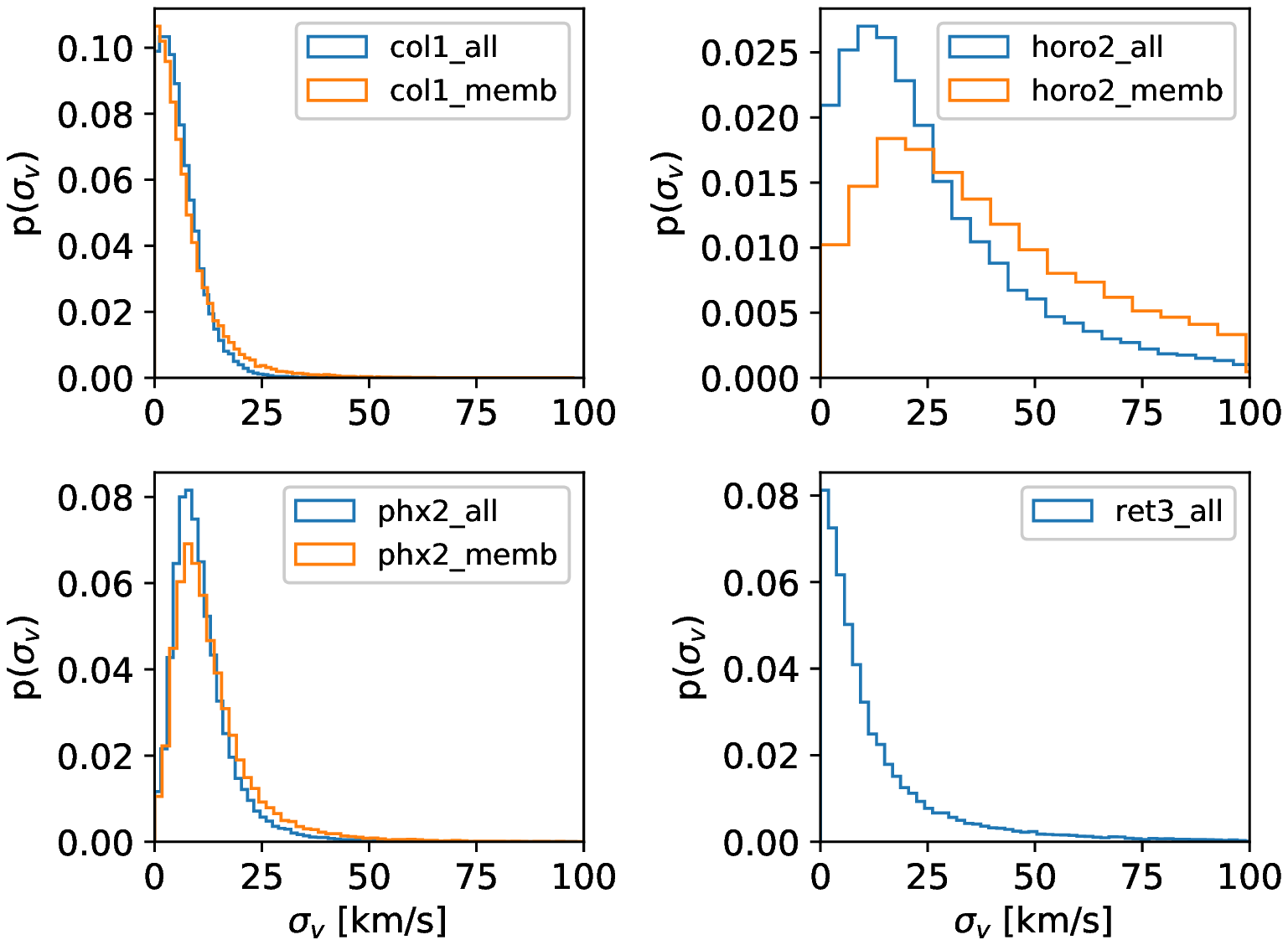}   
      \includegraphics[width=0.99\columnwidth]{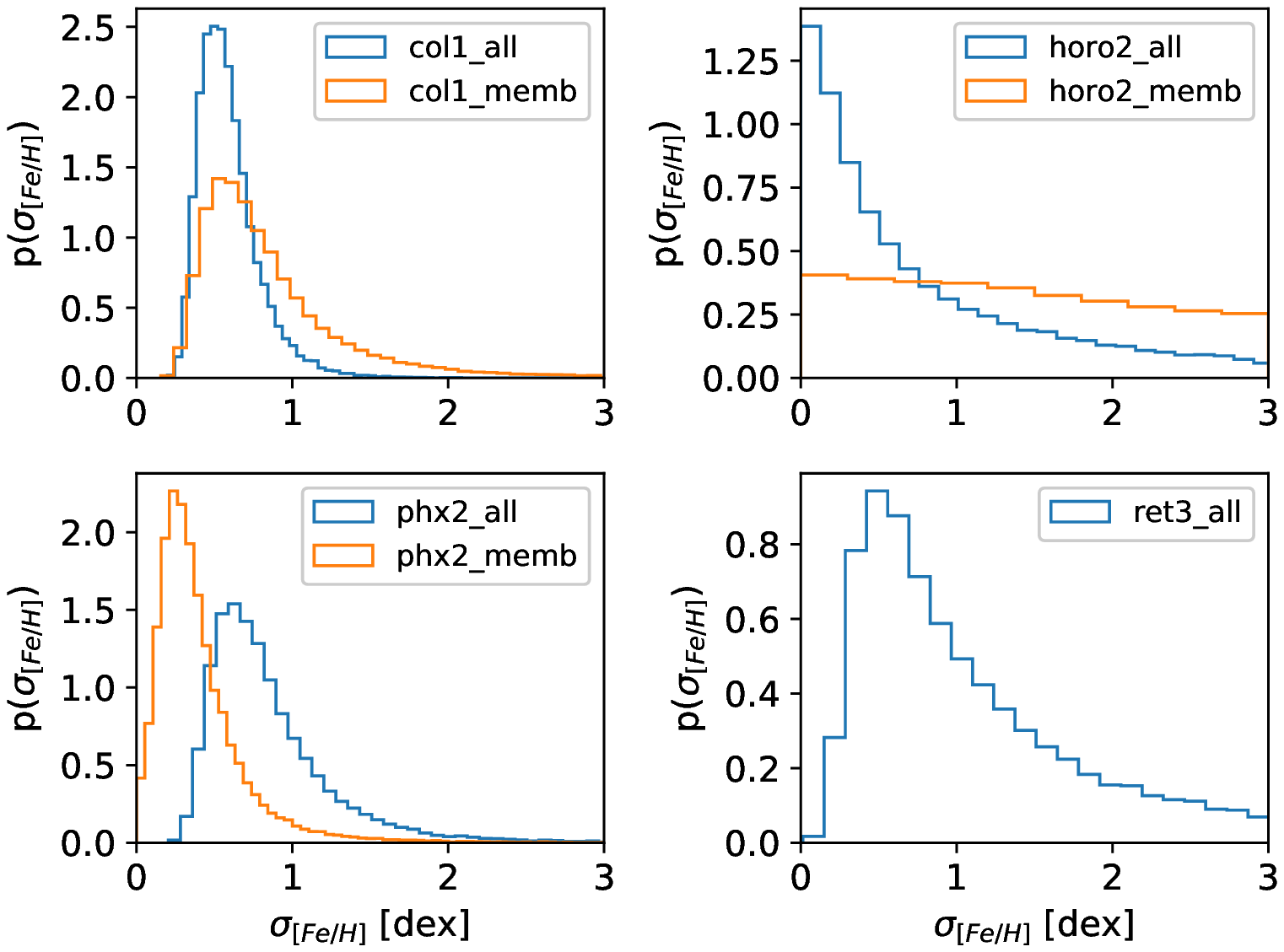}   
      \caption{Posterior probability distribution function of the dispersion in l.o.s. velocity (left) and metallicity (right) for the 4 satellites. The blue line shows the results when considering all potential members and the orange line only the most probable members.}
         \label{prob-dis}
   \end{figure*}

 Fig.~\ref{prob-dis} shows the posterior probability distribution functions (pPDF) obtained for $\sigma_{V, \mathrm{hel}}$ and $\sigma_{\mathrm{[Fe/H]}}$. In some cases the peak of the pPDF is found at a null spread, while in others the peak probability value is not very enhanced with respect to the probability at $\sigma_{V, \mathrm{hel}}=0$ (and $\sigma_{\mathrm{[Fe/H]}}=0$). Due to the importance of resolving the intrinsic spread in these systems to ascertain their nature, we complement our analysis with mock data-sets in order to understand whether the detection of an intrinsic spread might be occurring by chance, given the small number statistics and/or large measurement errors of the data-sets, or whether it is reliable. For each UF system, we produce 1000 mock samples of l.o.s. velocities and metallicities drawn from a Gaussian distribution with no intrinsic spread, which would mimic the case of the system being a stellar cluster, and with a size and error distribution as for the set of members being considered. As bin size we use everywhere a third of the lower 1$\,\sigma$ error in the measured parameters. For each of these mock samples, we measure the ratio of the probability at $\sigma_{V, \mathrm{hel}}=0$\,km/s (and $\sigma_{\mathrm{[Fe/H]}}=0$\,dex) over that of the probability at the peak of the PDF\footnote{With this definition, a ratio equal to zero implies a null probability at a dispersion equal to zero, while a ratio of one implies that the maximum of the PDF is found at a dispersion equal to zero.}: we consider that we are resolving the intrinsic spread when only 5\% of the mock sets shows a ratio as low as that given by the observations.

\begin{figure}
   \centering
   \includegraphics[width=0.72\columnwidth,angle=-90]{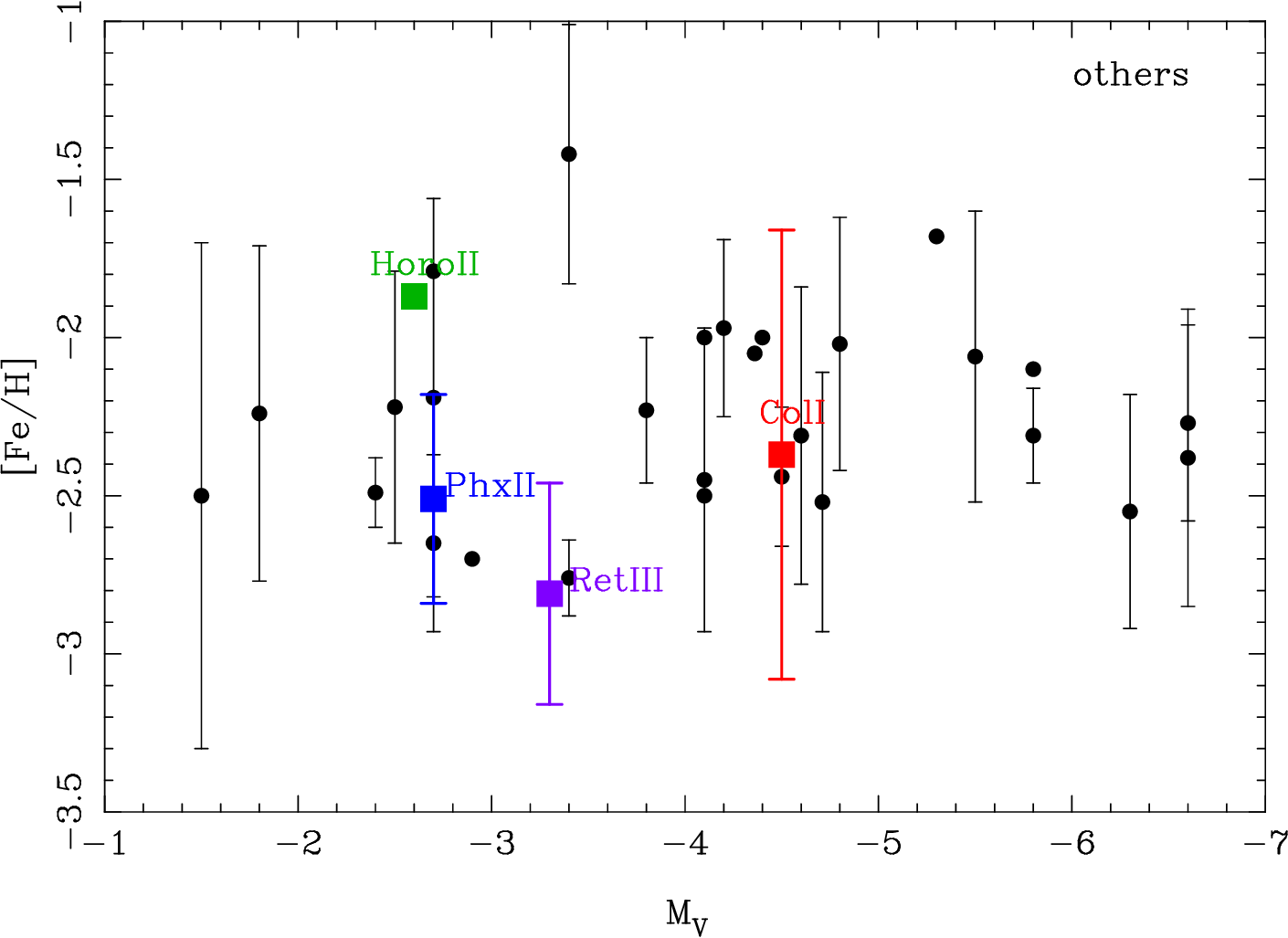}     
      \caption{Average metallicity and metallicity dispersion (the 'error' bars) as a function of absolute magnitude for the four target satellites. Objects that have no dispersion measurement or only limits have no error-bar plotted, like in case of Horo~II. The metallicities for the "other objects" are from \citet{Torrealba_16b,Martin_07,Simon_07,Koch_09,Carlin_09,Li_18a,Kirby_15,Li_17,Walker_16,Koposov_15b,Walker_09,Kim_16a,Simon_15,Simon_11,Kirby_13,Kirby_17,Simon_17,Li_18b,Koposov_18,Willman_11,Norris_10,Longeard_18}. The point at M$_V=-5.3$/[Fe/H]$=-1.68$ is Laevens~1/Crater~I which is probably a globular cluster.
              }
         \label{mass-met}
   \end{figure}

 While we resolve the intrinsic metallicity spread for all systems but Horologium~II, the only system for which the current data allow to resolve $\sigma_{V, \mathrm{hel}}$ is Phoenix~II. For the cases where the intrinsic spreads are unresolved, we report in Tab.~\ref{KapSou2} only upper limits (defined as the values at the 90-th percentile of the distributions). Figure~\ref{mass-met} shows where the objects fall onto the mass-metallicity plane, with respect to UF systems studied in the literature.

$\bullet$ {\it Impact of different samples of members:} We comment here on the set of values that we consider as the most reliable, among those derived from the samples of very likely members only and likely members + candidates. 

In the case of Phoenix~II, there is only one star with uncertain membership, which has a metallicity ([Fe/H] $= -1.1 \pm $0.18\,dex) significantly larger than the rest (see Tab.~2), thus our preferred set of values are those derived when excluding this star. The only parameter that is noticeably impacted by the choice of including/excluding this star is the spread in [Fe/H], which decreases from 0.75$_{-0.22}^{+0.43}$ dex to 0.33$_{-0.16}^{+0.29}$.

Also in the case of Col~I the candidate members are on average somewhat more metal rich than the rest; however, their metallicity values, [Fe/H]$ \leqslant-1.65$, are compatible with stars in dwarf galaxies of similar luminosity. To stay on the safe side, we consider as preferred values those obtained from the sample of very likely members only. We note that in general the properties of Col~I do not depend significantly on the sample chosen, since the candidates are faint (and thus have large errors on the derived properties) and have rather similar properties than the rest. In particular, there is a clear metallicity spread in this system in both cases.

For both Col~I and Phx~II, the proper motion values do not depend much on the sample used because the uncertain members have larger measurement errors and have therefore much less weight in the (weighted) average value. This is different for Horo~II, for which the uncertain member is bright. Given the difficulty of adopting a preferred sets of values for Horo~II, we retain both of them for the rest of the analysis.

$\bullet$ {\it Comparison with the literature:} Our independent measurement of the proper motion of Phx~II is in excellent agreement with that by K18, which was guided by the predictions from the model postulating an association to the LMC.

\citet{Pace_18} have determined proper motions for a set of UF objects, among which also the four systems studied here, without the aid of spectroscopic data-sets. In Appendix~\ref{ap_pace} we compare in detail with the results from their work. In summary, our preferred values differ by 1.6, 1.8, 0.9 and 1.9 $\sigma$ for Col~I, Hor~II, Phx~II and Ret~III\footnote{Note that in the calculation we adopt as error the largest one between that given in either of the two works, but do not combine them in quadrature, since the average proper motions are not fully independent.}. 
In all cases, the disagreement is small enough that it might have occurred by chance, although probably other reasons contribute, including an underestimate of the errors. Independently on this, as it will be clear in Sect.~\ref{sec:velocities}, apart from Phx~II, the current errors in systemic proper motions are large and further observations are necessary to obtain constrains on the orbit of these satellites.

$\bullet$ {\it Nature of the satellites} The only system for which we resolve the l.o.s. velocity dispersion is in Phx~II. Its value does not depend significantly on which set of members is considered and it is based on at least five stars that are partly so bright that a chance association is very unlikely. Even though a robust determination would benefit from larger sample sizes and multiple observations with different time sampling, it was shown that undetected binaries usually cannot inflate the measured l.o.s. velocity dispersion to the values measured for Phx~II \citep{McConnachie_10}. Phx~II intrinsic metallicity spread is 0.33\,dex in the case of the preferred set of members. We consider this value as rather robust because it contains the brightest (highest SNR) stars. Overall, the values of the l.o.s. velocity and metallicity dispersion for Phx~II are evidence that this satellite is a galaxy.

For Ret~III we measure a formally robust metallicity dispersion of 0.35 dex, but since the stars are rather faint,  we do not consider it as conclusive to classify the system as a galaxy.

The measurement for Col~I (0.71\,dex) is more robust since the two brightest certain members have a clearly different metallicity, [Fe/H]$=$-3.36 and -2.22.
The large metallicity spread appears to indicate that Col~I is a galaxy; this hypothesis is further supported by its rather large size \citep{Carlin_17}, that makes it unlikely to be a globular cluster.

 For the case of Horo~II our data are equally well consistent with a nature as globular cluster or galaxy, because our limits on intrinsic spreads are unconstraining.

We note however that,  even if some of the UFs are globulars clusters, it is likely that they originated within a dwarf galaxy, since halo globulars were very likely accreted \citep{Zinn_93,Muratov_10}. Thus, even in this case we might still learn about the orbital properties of their former hosts. We note that all our targets apart of maybe Horo~II have lower metallicities than most halo globulars, which is unlike Laevens~1 \citep{Weisz_16,Kirby_15} and Pyxis \citep{Fritz_17}. This could very tentatively suggest an origin in a smaller system, although it is known that globular clusters can be much more metal-poor than the field stars of their host (as in the case of the Fornax dSph).

      \begin{table*}
      \caption[]{Average and global properties of the objects in our sample. Column 1 lists the object name, col.~2 and 3 the average heliocentric velocity and its dispersion, col.~4 and 5 give the measured proper motion, col.~6 the average correlation coefficient between them, col.~7 and 8 the average metallicity and its dispersion. Above the line we show the values calculated from all possible members, below those from the certain members. We indicate in bold which determination we prefer (see main text for details). The upper limits refer to the 90-th percentile of the pPDFs.
      }
         \label{KapSou2}
       $
         \begin{array}{p{0.12\linewidth}llllllll}
            \hline
            satellite &  <V_{\rm hel}> [km/s] & \sigma_\mathrm{V,hel} [km/s] & <\mu_{\alpha^*}> [mas/yr] & <\mu_\delta> [mas/yr] & C_{\mu_{\alpha*},\mu_\delta} & <\mathrm{[Fe/H]}> &\sigma_{\rm [Fe/H]} \\
            \hline
            ColumbaI  & 148.7^{+3.7}_{-3.8}  & <12.2  & 0.27 \pm 0.22 & -0.48 \pm 0.30 & -0.16  & -2.14^{+0.20}_{-0.19}   & 0.55^{+0.21}_{-0.14} \\            
\textbf{Horologium~II}  & 168.7^{+12.9}_{-12.6} & <54.6 & 1.52 \pm 0.25 & -0.47 \pm 0.39 & 0.07  & -1.87^{+0.36}_{-0.50} & <1.93\\
Phoenix~II  & 32.6^{+4.7}_{-5.8}    & 9.5^{+6.8}_{-4.4} & 0.49 \pm 0.12 & -1.18 \pm 0.14 & -0.47  & -2.25^{+0.32}_{-0.33}  & 0.75^{+0.43}_{-0.22} \\
\textbf{Reticulum~III}  &  274.2^{+7.5}_{-7.4} & <31.2  & -0.39 \pm 0.53 & -0.32 \pm 0.63 & 0.45  & -2.81 \pm 0.29 & 0.35^{+0.21}_{-0.09} \\
\hline
\textbf{Columba~I}  & 153.7^{+5.0}_{-4.8}& <16.1 & 0.25 \pm 0.24 & -0.44 \pm 0.33 & -0.17  & -2.37^{+0.35}_{-0.34}   & 0.71^{+0.49}_{-0.24}  \\
Horologium~II  & 169.4^{+3.7}_{-3.8} & <75.9 & 0.36 \pm 0.68 & -0.5 \pm 1.08 & -0.17  & -2.10^{+1.02}_{-1.32} & <2.6\\

\textbf{Phoenix~II}  & 32.4^{+3.7}_{-3.8}  & 11.0^{+9.4}_{-5.3} & 0.5 \pm 0.12 & -1.16 \pm 0.14 & -0.47  & -2.51^{+0.19}_{-0.17}  & 0.33^{+0.29}_{-0.16}  \\
     \noalign{\smallskip}
            \hline
         \end{array}
   $  
   \end{table*}

\section{Velocities, orbital poles and association to the LMC}  
\label{sec:velocities}

\begin{figure}
   \centering
   \includegraphics[width=0.72\columnwidth,angle=-90]{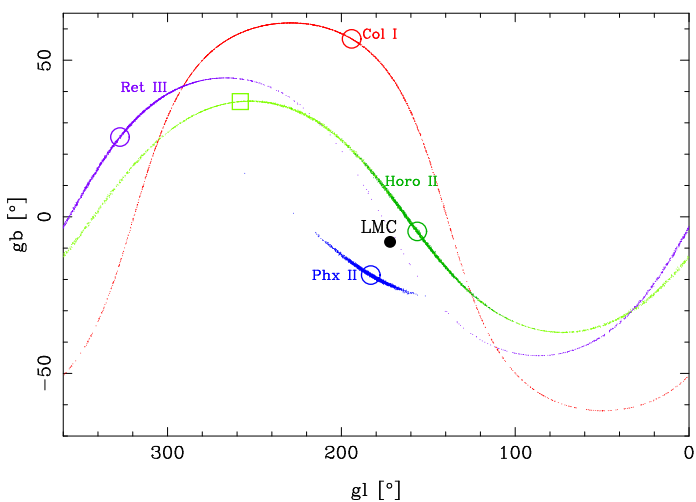}  
      \caption{All-sky view of orbital poles for the objects in the sample; the open circles (or square) indicate the value which follows from the measurements while the small points around each object plot the orbital poles from the individual 
Monte Carlo simulations.  We show for all satellites our preferred membership selection; for Horo~II, we also show the other option in light green, which covers nearly the full great circle. For the other satellites the two options lead to very similar results.}
         \label{poles}
   \end{figure}

We now use the systemic proper motions and l.o.s. velocities derived in the previous section to determine the Galactocentric velocities and orbital parameters of the systems in the sample (see Table~\ref{KapSou3} and Fig.~\ref{poles}). 

For the conversion into velocities, and for the following analysis, we also add a systematic error which depends on the size of the system to the proper motions errors in both dimensions \citep[see][for the details]{Fritz_18b}, and we assume it to be uncorrelated between $\alpha$ and $\delta$. Given the large random errors for our stars, the systematics are not important. We followed the same procedure as \citet{Fritz_18b} to convert the observed heliocentric velocities and proper motions into Galactocentric properties. This also accounts for uncertainties into the solar motion and distance of the satellite, but in our cases the total uncertainties are dominated by the proper motion uncertainties. As discussed in \citet{Fritz_18b}, in this situation classical forward Monte Carlo simulations are still appropriate for orbital poles estimates, but backward Monte Carlo simulations are to be preferred to avoid biases in positive defined quantities such as the tangential velocity, V$_\mathrm{tan}$.

  \begin{table}
      \caption[]{Galactocentric distances and velocities for the satellites in our sample. Col.~1 shows the name, col.~2 the object distance, col.~3, 4 and 5 give the radial, tangential and 3D velocity, respectively. In the top part we provide the determinations from the full sample of members, while in the bottom part using the very likely members. We indicate in boldface our preferred value.} 
         \label{KapSou3}
       $$
         \begin{array}{p{0.15\linewidth}lllll}
            \hline
            satellite &  d_{\rm GC} [kpc] & V_\mathrm{rad} [km/s] & V_\mathrm{tan} [km/s] & V_\mathrm{3D} [km/s] \\
            \hline
Col I  & 185 & -27\pm4 & 308^{+186}_{-114} & 309^{+187}_{-113} \\
\textbf{Horo II} & 76 & 35\pm9 & 378^{+133}_{-124} & 380^{+133}_{-124} \\
Phx II  & 80 & -42\pm3 & 257^{+69}_{-67} & 260^{+69}_{-65} \\
\textbf{Ret III}  & 100 & 92\pm9 & 296^{+249}_{-200} & 310^{+241}_{-174} \\
\hline
\textbf{Col I}  & 185 & -24\pm4 & 224^{+235}_{-160} & 226^{+234}_{-157} \\
Hor II  & 76 & 19\pm7 & 96^{+101}_{-65} & 98^{+100}_{-60} \\
\textbf{Phx II}  & 80 &-42\pm4 & 255^{+65}_{-66} & 258^{+64}_{-65} \\

    \noalign{\smallskip}
           \hline
       \end{array}
       $$
  \end{table}

 Figure~\ref{poles} shows the estimated orbital poles with their uncertainties, compared to the orbital pole of the LMC. Phoenix~II has the best localized pole direction:  $l=185\pm9$ and $b=-18\pm3$ degree, which is at 16\degree~from the pole of the LMC and  close to the location of the pole of Hydra~I, a galaxy considered to be a former LMC satellite in K18. In the other cases the error in proper motion is large, making nearly a great circle possible.  It appears however unlikely that Col~I was once associated to the LMC as the direction of its orbital pole is always at least 32\degree~from the LMC one. 
 An association of Horo~II to the LMC becomes significantly more or less likely depending on whether the brightest (most distant in R$_\mathrm{half}$) candidate member is truly a member or not. When this star is included, the most likely pole is formally determined with a higher accuracy, $l=156\pm17$ and $b=-5\pm12$ degree, and it is at only 16 degree distance of the LMC. When excluding it, the most likely pole direction changes very significantly (to l$=258$ and b$=37$) and the 1-$\sigma$ region spans nearly two thirds of the sky, making a prior association with the LMC less likely (although not completely excluded). The inclusion/exclusion of this star changes quite dramatically also the values of the tangential and 3D velocity.
Ret~III is still an uncertain case, since the LMC can be found within 3\degree~from the great circle allowed by the Monte Carlo simulations, but only in a minority of cases.

For most galaxies the total velocity is very high but the errors are too large for the measurement to be informative. The exception is Phx~II (V$_\mathrm{3D}=258\pm65$\,km/s); its orbit would have a eccentricity of 0.76$^{+0.13}_{-0.40}$ (0.33$^{+0.27}_{-0.16}$) in a low (high) mass Milky Way halo\footnote{See \citet{Fritz_18b} for the definition of the halos.}; therefore Phx~II might have been accreated on a high angular momentum orbit, as the LMC. 
 
 We note in passing that the proper motion for Ret~III by \citet{Pace_18} corresponds to a bias corrected $v_{\rm 3D}=750^{+210}_{-182}$ km/s with a pole similar to our determination but better constrained, thus fully inconsistent with the LMC. 
If real this velocity would be very interesting, since it is more than 1 $\sigma$ larger than a generous escape speed of the Milky Way\footnote{For the escape velocity we use a MW halo of $1.6\times 10^{12}$ M$_\odot$ as in \citet{Fritz_18b}.}. 

The determinations of the systemic proper motions and l.o.s. velocities, combined to the satellites' distance, give a  further  opportunity of exploring a possible origin as former LMC satellites by comparing to the  predictions given by \citet{Jethwa_16} and K18 for these observables, which also avoids the issues of error propagation onto the Galactocentric velocities. 

For Horo~II, in the case of the generous membership sample, our measurements match very well the predictions by K18 but not those by \citet{Jethwa_16}. On the other hand, the estimate from the sample of very likely members is compatible within 1-2\,$\sigma$ with both studies, given the larger error-bars. 

Ret~III matches the predictions by \citet{Jethwa_16} in all quantities within 1-2\,$\sigma$ both for an association to the LMC or to the SMC. It also agrees well with the model in K18 in $\mu_{\delta}$ and $v_{\rm helio}$, but not in $\mu_{\alpha\,*}$.

For Col~I the agreement is good in the proper motions with both models (mostly caused by the large  errors of the observations) but entirely off in l.o.s. velocity for K18 and compatible with \citet{Jethwa_16} within 2\,$\sigma$. We point out though that the typical error-bars in the l.o.s. heliocentric systemic velocity by the latter authors are very large for all satellites, at least $\pm$70-80\,km/s. Since Col~I is located outside the main track of LMC debris and has an unfitting large distance for its location, we deem it as very unlikely that it is a former member of the LMC cohort of satellites. 

 As mentioned in the previous section, we recover closely the proper motion of Phx~II deduced by K18\footnote{For this galaxy the authors guide their systemic proper motion determination by searching for a clump in proper motion space in the region predicted by the model.}; however, our systemic l.o.s. velocity is off by 45 km/s, about 9\,$\sigma$ away from their prediction, which is $-15.5_{-10.5}^{+5.2}$ (here, $\sigma$ is the error in the predicted l.o.s. velocity). Our proper motion measurements do not agree with \citet{Jethwa_16} model predictions.  The problems in K18 to predict the properties exactly could be associated with the fact that they use a light-LMC analog (see e.g.  Sect.~\ref{sec:lmcmass}). \citet{Jethwa_16} consider a range of usually more massive analogs, which result in the predicted properties encompassing a much larger range of values than in K18. Still, the ranges given by \citet{Jethwa_16} and K18 differ, especially for the proper motions. There are several possible reasons for that. It might be that the difference in the LMC mass is the culprit, but it is also possible that other more subtle effects are responsible like differences in the conversion into the solar reference frame.

 Given the close orbital pole of Phx~II to that of the LMC and the good agreement with the K18 predictions in most properties, we conclude that Phx~II is the only object for which the case of a former association to the LMC can be made. Given the similarity of the LMC and Vast Polar Structure orbital pole, this also implies that Phx~II could be part of this structure \citep[e.g.][]{Pawlowski_12}.

\section{On the LMC mass}\label{sec:lmcmass}
At present there are 42 galaxies close to the Milky Way that have sufficiently good proper motions and l.o.s. velocities to classify them as either LMC or MW satellite (all galaxies in Fritz et al. \citeyear{Fritz_18b} apart from Phoenix~I, the SMC, Phoenix~II and Col~I from this work, Antlia~II from Torrealba et al. \citeyear{Torrealba_18b}). The properties of five to eight UFDs are compatible with them having formed part of the satellite system of the LMC: the SMC \citep[see e.g.][]{DOnghia_16}, Hydrus~I, Carina~II, Carina~III, Horologium~I \citep{Kallivayalil_18} and, possibly, Phoenix~II (this work),  Draco~II and Hydra~II \citep{Kallivayalil_18}.

In principle the number of LMC satellites could be translated into a mass ratio of the LMC versus the Milky Way, and then into a mass for the LMC, assuming the MW mass is known. In practise, there are a number of caveats. First of all, 
one needs to assume how the number of luminous satellites scales with the mass of the host, something that it is not well-known in the regime of hosts with low mass and that carries with it the uncertainties related to the formation of a luminous components in the smallest sub-haloes. One should also consider that the number of known satellites for both systems is incomplete to some extent. On the one hand, the completeness is likely slightly lower for the Milky Way, since the Galactic plane obscures less of the Magellanic system and because fewer of the Magellanic satellites should be at very larger distance from the Sun, because less massive halos are less extended. On the other hand the incomplete spectroscopy of recently discovered satellites acts in the opposite direction; to what extent the two effects compensate is open to question. Finally, the mass of the Milky Way itself is still subject of debate, with values varying up to a factor of 2.

Let us assume that the number of luminous satellite galaxies scales with the host halo mass as the sub-haloes in \citet{Gao_04} and \citet{Wetzel_15}. We then consider four options, varying the number of LMC satellites, $N_{\rm LMC}$, between 5 and 8 (and the number of MW satellites as 42 - $N_{\rm LMC}$), and adopt a Gaussian error given by the square-root of the number of satellites. 
This yields a median mass ratio of $0.18^{-0.08}_{+0.09}$, similar to that obtained by \citet{Penarrubia_16} from the timing argument (0.2) and by \citet{Erkal_18b} using the Orphan stream (0.15). If we take $1.1\pm0.3\times10^{12}$\,M$_{\odot}$ as M$_\textrm{200}$  of the Milky Way \citep{Bland_16} and assume that the mass ratio and  the MW mass are independent of each other, we obtain a  M$_\textrm{200}$ mass for the LMC of  $1.9_{-0.9}^{+1.3}\times10^{11}$\,M$_\odot$. 

The error-bar on our estimate of the LMC mass is large, but at least from the point of view of reasonable values for the MW mass, it 
seems unlikely that the LMC $M_\textrm{200}$ was as low as $3.6\times10^{10}$\,M$_{\odot}$ before infall as the analog used in \citet{Kallivayalil_18}. Qualitatively, also the fact that potential satellites like Phx~II and Hydrus~II are found at large distances from the LMC points also in a similar direction. More models exploring LMC-analogs of high mass are necessary to test whether these would produce tidal debris spread at large distances similar to those of Draco~II, Hydra~II and Phoenix~II. 

Finally, we note that the increasing number of satellites around the Milky Way that are being classified as LMC former satellites has the effect of decreasing the number of dwarf galaxies that are considered as luminous satellites of the Milky Way.  

 \section{Summary and conclusions}
 \label{sec:summ}
 We present results from FLAMES/GIRAFFE intermediate resolution spectroscopy in the region 8200-9200\AA\, for individual stars in the l.o.s. to four of the newly discovered Milky Way satellites that are lacking spectroscopic information in the literature: Columba~I, Horologium~II, Phoenix~II and Reticulum~III. This implies that their nature as globular clusters or dwarf galaxies was not established and even basic properties such as their systemic l.o.s. velocity were unknown. Horo~II and Ret~III also lack determinations of their systemic proper motion, while for Col~I and Phx~II the only existing determination was obtained by K18 searching the proper motion space assuming that these objects would have been prior satellites of the LMC, guided by a ``light LMC'' N-body model. 

We extracted the l.o.s. velocity and metallicity of the target stars via a spectral fitting analysis, assuming photometric estimates of the stars's effective temperature and gravity. The information coming from the spectroscopy is used together with GDR2 astrometric data and photometry (also from the DECam NSC DR1 catalog) to search for stars that are probable members to these systems. All the velocity peaks that we associate to detection of member stars in these systems are found to be highly unlikely to be due to polluter stars passing our selection criteria. 

Once the systemic velocity of the system is known, additional members are looked for among the stars targeted by FLAMES/GIRAFFE but without GDR2 kinematic informations. In total we find 15 very likely members (5 in Col~I, 2 in Horo~II, 5 in Phx~II, 3 in Ret~III) at most within 2.1 R$_\mathrm{half}$. In addition we have 6 candidate member stars (4 for Col~I, 1 for Horo~II and 1 for Phx~II), usually at larger projected radii. 
 
The system with the most constrained properties is Phx~II; for this object, we resolve its l.o.s. velocity dispersion ( $11.0_{-5.3}^{+9.4}\,$km/s), for the preferred membership selection; we also find an intrinsic metallicity spread of at least 0.33$_{-0.16}^{+0.29}\,$dex; these properties suggest its nature as a dwarf galaxy. Col~I has also a robust metallicity spread, since the two brightest members have a metallicity difference of 1.1\,dex. 
The results  on the nature of Horo~II and Ret~III are inconclusive, as they are more sensitive to the adopted sample, and we cannot robustly place the objects in one category or another one. Nonetheless, should these objects be globular clusters, being in the outer halo, they might have formed within an accreted dwarf system and therefore still provide information on their former host, such as e.g. its orbital properties.

Along the great circle track allowed by the measurement errors, Col~I orbital pole remains always at least 32\degree~from the LMC, making an association unlikely. On the other hand, we cannot firmly exclude an association between the LMC and Horo~II or Ret~III, although the latter appears rather unlikely due to the small fraction of Monte Carlo realizations that could bring its orbital pole close to the one of the LMC. 

 The orbital pole of Phx~II is about 16\degree~away from the one of the LMC, and its  systemic proper motion agrees well with the predictions by \citet{Kallivayalil_18}, although the l.o.s. systemic velocity is 9$\,\sigma$ away. However, it is possible that a LMC-analog producing a larger spread in the tidal debris properties would lead to a better agreement also in this observable. Therefore Phx~II appears as a promising system for having been a former LMC satellite. Given the similarity of the LMC and Vast Polar Structure orbital pole, this also implies that Phx~II could be part of this structure \citep[e.g.][]{Pawlowski_12}. Bar the caveats discussed, the number of potential LMC former satellites (Phx~II and those from works in the literature) would suggest a large mass before infall for the LMC, of the order of $2\times10^{11} M_{\odot}$.

\begin{acknowledgements}
This work has made use of data from the European Space Agency (ESA) mission
{\it Gaia} (\url{https://www.cosmos.esa.int/gaia}), processed by the {\it Gaia}
Data Processing and Analysis Consortium (DPAC,
\url{https://www.cosmos.esa.int/web/gaia/dpac/consortium}). Funding for the DPAC
has been provided by national institutions, in particular the institutions
participating in the {\it Gaia} Multilateral Agreement.
      
G.B. gratefully acknowledges financial support by the Spanish Ministry of Economy and Competitiveness (MINECO) under the Ramon y Cajal Programme (RYC-2012-11537); G.B. and S.T. both acknowledge support by MINECO under the grants AYA2014-56795-P and AYA2017-89076-P.

\end{acknowledgements}
\begin{appendix}
\label{sec:app}
\section{Comparison with Pace and Li}
\label{ap_pace}
\citet{Pace_18} have determined systemic proper motions for UF systems in the DES with a method that uses photometric information from the first public data release of DES data and proper motions from GDR2. The set of objects include the four satellites analyzed in this work and, as mentioned in the main text, we differ in our mean proper motions by 1.6, 1.8, 0.9 and 1.9 $\sigma$ for Col~I, Hor~II, Phx~II, Ret~III, respectively.

To find the reason for this small disagreement we first check the membership classification on a star by star basis. 

In total 26 of our stars with measured l.o.s. velocities are in the \citet{Pace_18} catalog (hereafter, PL18). Among these, all the stars that we have classified as highly likely members have at least a membership probability $p$ of 82\% in PL18 catalog (apart from the BHB candidate in Hor~II, which has $p=$68\%). Among our candidate members, we found matches only for Col~I, and these have $p$ between 10\% and 70\%. 

Seven of our stars are missing from the PL18 list, of which four with measured proper motions (hor2\_2\_48, phx2\_8\_24, phx2\_5\_46, ret3\_2\_70). Not surprisingly, that includes a larger fraction of candidate members. phx2\_8\_24, phx2\_5\_46 and ret3\_2\_70 are possibly missing either because too red for the PL18 selection or because of their red-HB-like location on the CMD, a region that was excluded by the authors; hor2\_2\_48 is likely missing due its large distance from the satellite center. We had included hor2\_2\_48 in the set of members giving our preferred value, since it is so bright that an interloper origin is less likely. Due to its large spectroscopic metallicity,  phx2\_8\_24 instead was not part of the sample of members used for the preferred set of values. phx2\_5\_46 is bright and its properties agree well with those of the other members in the three components of the motion, thus very likely it belongs to Phx~II.

Of our non-members only one has more than 10\% probability of membership in PL18 catalog  (col1\_8\_1,5 with 77\%). 

From the PL18 members there are 2, 3, 5 and 3 (for the four satellites) stars with constraining properties ($p>$50\%, the majority of which with $p>$90\% and $\delta \mu<$ 4 mas/yr 
in \citet{Pace_18} for which we have no l.o.s. velocity information. All but one of these stars would have passed our selection criteria, however either there was no FLAMES/GIRAFFE spectrum available for those stars or their l.o.s. velocity determination was unreliable, and they were therefore excluded from our analysis.

In the following we discuss how the systemic motion would be impacted by the inclusion/exclusion of stars to one of the two samples. If col1\_8\_15 ($v_\mathrm{helio}=82$ km/s) would be excluded from the PL18 sample, their proper motion determination would change towards our values, but not enough to make it compatible within 1$\sigma$. We find that the main difference for Col~I is caused by one bright stars at ($\alpha$, $\delta$)$=$ (49.1666, $-$50.0469) that is among the PL18 members but that was not targetted in the FLAMES/GIRAFFE observations.
For Hor~II we give only a slight preference to our preferred membership selection; the other choice leads to agreement in proper motion within 1$\,\sigma$ but at the cost of a clearly larger error. In case of Phx~II the agreement was already good. 
When we exclude ret3\_2\_70, our motion still disagrees to 1.6 $\sigma$ with the motion of \citet{Pace_18}. Such a difference occurs with a probability of 26\%. 

Thus overall, we do not find clear-cut reasons as to why we slightly disagree. Since the disagreement is small it might have occurred by chance. An underestimate of the errors could be another possible cause: e.g. we treat the stars with a binary membership classification rather than deriving the properties factoring in the probability of membership. PL18 do provide continuous probabilities, however the CMD information is treated in a binary way. Also, Pl18 spatial cut is rather close to the center of the satellite, which might exclude possible tidally affected stars.

Finally, since both studies adopt a photometric pre-selection of targets but on different catalogues (NSC DR1 versus DES DR1), we used the stars in common to test for systematic shifts that could impact on the selection and found that these were very small (0.02mag in g/r/i-band).
Overall, the stars only identified in  PL18 tend to be bluer than the stars for which we have l.o.s. velocity information. Partly, that is due to increased difficulties to get l.o.s. velocities for very blue stars, partly due to a possible too restrictive (on the red side) selection box in PL18. 

Independent of which photometric catalogue is used both our members and the members of PL18 span a bigger range in color (also reaching the location of a [Fe/H]$=$-1 isocrone) than expected for red giant branch stars on the basis of the measured spectroscopic metallicities and of the rareness of metal-rich stars in ultra faint galaxies. 

\begin{figure}
   \centering
   \includegraphics[width=0.72\columnwidth,angle=-90]{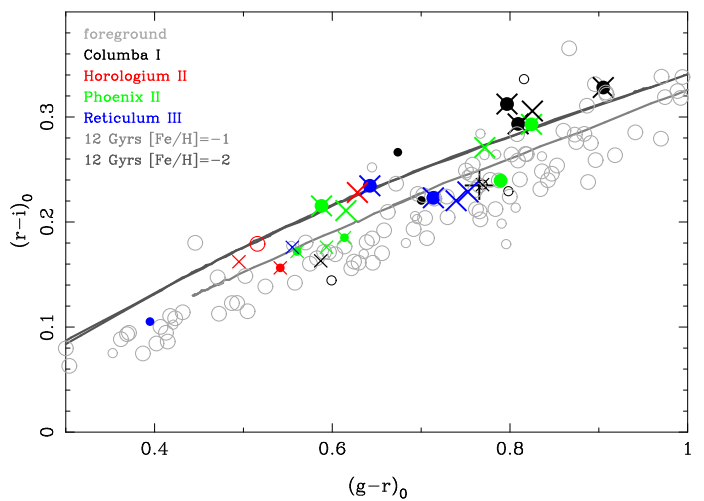}  
      \caption{Color-color diagram of all our (potential) members for metallicity estimation following PL18 (filled circles: highly likely members; open circles: candidates). The members of \citet{Pace_18} ($p >0.5$) are indicated by crosses. The member of \citet{Pace_18} not confirmed by spectroscopy is indicated with a plus.  Foreground stars are selected by requiring that they do not pass our halo selection based on \textit{Gaia} kinematics. Larger symbols are stars brighter than m$_i=20$. The Parsec isochrones used have only evolutionary stages brighter than the turnoff. 
              }
         \label{color-color}
   \end{figure}

On the other hand, we confirm the finding of PL18 that the g-r/r-i color-color diagram can select metal poor stars, see Figure~\ref{color-color}. In our case it only works for stars brighter than m$_i=20$, probably due to the increasingly larger errors in the colors. The majority of our members are above the foreground stars sequence as in \citet{Pace_18} and close to the [Fe/H]$=-2$ curve. 
Thus, we confirm that the location of metal-poor stars on the color-color diagram is more according to expectations than that on the color magnitude diagram. The reason for that is unclear.

As PL18, we find that the location of Ret~III members on the color-color plot would suggest a larger metallicity than for the rest of the UF systems analyzed (even though the spectroscopic metallicities here derived do not back up the suggestion from the color-color plot). 

\end{appendix}
\bibliographystyle{aa} 
\bibliography{mspap} 
\end{document}